\documentclass[twocolumn]{aastex631}
\usepackage{xspace}

\newcommand\aastex{AAS\TeX}

\newcommand{\JWST}{{\em JWST}\xspace}

\shorttitle{}
\shortauthors{Vulcani et al.}
\graphicspath{{./}{figures/}}

\begin{document}

\title{Early results from  GLASS-JWST. XX: Unveiling a population of ``red-excess'' galaxies in  Abell2744 and in the coeval field
}

\correspondingauthor{Benedetta Vulcani}
\email{benedetta.vulcani@inaf.it}

\author[0000-0003-0980-1499]{Benedetta Vulcani}
\affiliation{INAF- Osservatorio astronomico di Padova, Vicolo Osservatorio 5, I-35122 Padova, Italy} 

\author[0000-0002-8460-0390]{Tommaso Treu}
\affiliation{Department of Physics and Astronomy, University of California, Los Angeles, 430 Portola Plaza, Los Angeles, CA 90095, USA}

\author[0000-0003-2536-1614]{Antonello Calabr\`o}
\affiliation{INAF - Osservatorio Astronomico di Roma, via di Frascati 33, 00078 Monte Porzio Catone, Italy}

\author[0000-0002-7042-1965]{Jacopo Fritz}
\affiliation{Instituto de Radioastronomia y Astrofisica, UNAM, Campus Morelia, AP 3-72, CP 58089, Mexico}

\author[0000-0001-8751-8360]{Bianca M. Poggianti}
\affiliation{INAF- Osservatorio astronomico di Padova, Vicolo Osservatorio 5, I-35122 Padova, Italy}

\author[0000-0003-1383-9414]{Pietro Bergamini}
\affiliation{Dipartimento di Fisica, Università degli Studi di Milano, via Celoria 16, I-20133 Milano, Italy}
\affiliation{INAF - OAS, Osservatorio di Astrofisica e Scienza dello Spazio di Bologna, via Gobetti 93/3, I-40129 Bologna, Italy}

\author{Andrea Bonchi}
\affiliation{Space Science Data Center, Italian Space Agency, via del Politecnico, 00133, Roma, Italy}

\author[0000-0003-4109-304X]{Kristan Boyett}
\affiliation{School of Physics, University of Melbourne, Parkville 3010, VIC, Australia}
\affiliation{ARC Centre of Excellence for All Sky Astrophysics in 3 Dimensions (ASTRO 3D), Australia}

\author[0000-0001-6052-3274]{Gabriel B. Caminha}
\affiliation{Technische Universität München, Physik-Department, James-Franck Str. 1, 85748 Garching, Germany}
\affiliation{Max-Planck-Institut f\"ur Astrophysik, Karl-Schwarzschild-Str. 1, D-85748 Garching, Germany}

\author[0000-0001-9875-8263]{Marco Castellano}
\affiliation{INAF - Osservatorio Astronomico di Roma, via di Frascati 33, 00078 Monte Porzio Catone, Italy}

\author[0000-0002-6317-0037]{Alan~Dressler}
\affiliation{The Observatories, The Carnegie Institution for Science, 813 Santa Barbara St., Pasadena, CA 91101, USA}

\author[0000-0003-3820-2823]{Adriano Fontana}
\affiliation{INAF - Osservatorio Astronomico di Roma, via di Frascati 33, 00078 Monte Porzio Catone, Italy}

\author[0000-0002-3254-9044]{Karl Glazebrook}
\affiliation{Centre for Astrophysics and Supercomputing, Swinburne University of Technology, PO Box 218, Hawthorn, VIC 3122, Australia}

\author[0000-0002-5926-7143]{Claudio Grillo}
\affiliation{Dipartimento di Fisica, Università degli Studi di Milano, via Celoria 16, I-20133 Milano, Italy}
 \affiliation{INAF—IASF Milano, via A. Corti 12, I-20133 Milano, Italy}

\author[0000-0001-6919-1237]{Matthew A. Malkan}
\affiliation{Department of Physics and Astronomy, University of California, Los Angeles, 430 Portola Plaza, Los Angeles, CA 90095, USA}

\author[0000-0002-9572-7813]{Sara Mascia}
\affiliation{INAF - Osservatorio Astronomico di Roma, via di Frascati 33, 00078 Monte Porzio Catone, Italy}

\author[0000-0001-9261-7849]{Amata Mercurio}
\affiliation{Dipartimento di Fisica “E.R. Caianiello”, Università Degli Studi di Salerno, Via Giovanni Paolo II, I–84084 Fisciano (SA), Italy}
\affiliation{INAF -- Osservatorio Astronomico di Capodimonte, Via Moiariello 16, I-80131 Napoli, Italy}

\author[0000-0001-6870-8900]{Emiliano Merlin}
\affiliation{INAF - Osservatorio Astronomico di Roma, via di Frascati 33, 00078 Monte Porzio Catone, Italy}

\author[0000-0002-8632-6049]{Benjamin Metha}
\affiliation{Department of Physics and Astronomy, University of California, Los Angeles, 430 Portola Plaza, Los Angeles, CA 90095, USA}
\affiliation{School of Physics, University of Melbourne, Parkville 3010, VIC, Australia}
\affiliation{ARC Centre of Excellence for All Sky Astrophysics in 3 Dimensions (ASTRO 3D), Australia}

\author[0000-0002-8512-1404]{Takahiro Morishita}
\affiliation{IPAC, California Institute of Technology, MC 314-6, 1200 E. California Boulevard, Pasadena, CA 91125, USA}

\author[0000-0003-2804-0648 ]{Themiya Nanayakkara}
\affiliation{Centre for Astrophysics and Supercomputing, Swinburne University of Technology, PO Box 218, Hawthorn, VIC 3122, Australia}

\author[0000-0002-7409-8114]{Diego Paris}
\affiliation{INAF - Osservatorio Astronomico di Roma, via di Frascati 33, 00078 Monte Porzio Catone, Italy}

\author[0000-0002-4140-1367]{Guido Roberts-Borsani}
\affiliation{Department of Physics and Astronomy, University of California, Los Angeles, 430 Portola Plaza, Los Angeles, CA 90095, USA}

\author[0000-0002-6813-0632]{Piero Rosati}
\affiliation{Dipartimento di Fisica e Scienze della Terra, Università degli Studi di Ferrara, Via Saragat 1, I-44122 Ferrara, Italy}
\affiliation{INAF - OAS, Osservatorio di Astrofisica e Scienza dello Spazio di Bologna, via Gobetti 93/3, I-40129 Bologna, Italy}

\author[0000-0002-4430-8846]{Namrata Roy}
\affiliation{Center for Astrophysical Sciences, Department of Physics and Astronomy, Johns Hopkins University, Baltimore, MD, 21218}

\author[0000-0002-9334-8705]{Paola Santini}
\affiliation{INAF - Osservatorio Astronomico di Roma, via di Frascati 33, 00078 Monte Porzio Catone, Italy}

\author[0000-0001-9391-305X]{Michele Trenti}
\affiliation{School of Physics, University of Melbourne, Parkville 3010, VIC, Australia}
\affiliation{ARC Centre of Excellence for All Sky Astrophysics in 3 Dimensions (ASTRO 3D), Australia}

\author[0000-0002-5057-135X]{Eros Vanzella}
\affiliation{INAF -- OAS, Osservatorio di Astrofisica e Scienza dello Spazio di Bologna, via Gobetti 93/3, I-40129 Bologna, Italy}

\author[0000-0002-9373-3865]{Xin Wang}
\affil{School of Astronomy and Space Science, University of Chinese Academy of Sciences (UCAS), Beijing 100049, China}
\affil{National Astronomical Observatories, Chinese Academy of Sciences, Beijing 100101, China}
\affil{Institute for Frontiers in Astronomy and Astrophysics, Beijing Normal University,  Beijing 102206, China}

\begin{abstract}
We combine   JWST/NIRCam imaging and MUSE data to characterize the properties of galaxies in different environmental conditions in the cluster Abell2744 ($z=0.3064$) and in its immediate surroundings. {We investigate how galaxy colors, morphology and star forming fractions depend on wavelength and on different parameterizations of environment.} Our most striking result is the discovery of a ``red-excess'' population in F200W$-$F444W colors  both in the cluster regions and the field. These galaxies have normal F115W$-$F150W colors, but are up to 0.8 mag redder than red sequence galaxies in F200W$-$F444W. They also have rather blue rest frame B$-$V colors. {Galaxies in the field and at the cluster virial radius are overall characterized by redder colors, but galaxies with the largest color deviations are found in the field and in the cluster core. Several results}  suggest that mechanisms taking place in these regions might be more effective in producing these colors. Looking at their morphology, many cluster  galaxies show signatures consistent with ram pressure stripping, while field galaxies have features resembling interactions and mergers. Our hypothesis is that these galaxies are characterized by dust enshrouded star formation: a JWST/NIRSpec spectrum for one of the galaxies is dominated by a strong PAH at 3.3$\mu m$, suggestive of dust obscured star formation. Larger spectroscopic samples are needed to understand if the color excess is due exclusively to dust-obscured star formation, and the role of environment in triggering it.

\end{abstract}


\section{Introduction} \label{sec:intro}
It is now widely accepted that galaxy clusters, the densest associations of matter in the Universe, are able to impact the properties of their galaxies. The  physical properties of cluster galaxies are found to be starkly different from those of their field counterparts, in terms of colors, star formation activity, morphologies, structural parameters and ages \citep[e.g.,][]{Dressler1980, Postman1984, dressler97, hashimoto98, lewis02, balogh04a, postman05, mastropietro05, vonderlinden10, Sheen2017, Song2017, Vulcani2010ComparingEnvironments, Paccagnella2016, Vulcani2017The0.7, Vulcani2018_L, Webb2020}.

Differences emerge because dense environments are believed to accelerate galaxy evolution, producing more massive systems faster \citep[e.g.,][]{delucia04c, gao05, Maulbetsch2007, Shankar2013, Papovich2012, Rudnick2012, Andreon2013, Newman2014, morishita16}. The direct role of the cluster environment in driving galaxy evolution through cluster specific effects, however, is still largely unclear.
Various physical mechanisms, which are either gravitational or hydrodynamic in nature, are often invoked. While the gravitational interactions mainly culminate in the development of tidal features, the hydrodynamic interactions between the galactic interstellar medium (ISM) and the hot intracluster medium (ICM) may lead to gas removal and eventually quench star formation \citep[see, e.g.,][for a review]{Boselli2006}.

To complicate the picture, clusters are a very heterogeneous population, spanning a wide range of sizes, densities and temperatures of the ambient medium, and in different phases of their evolution. Some of them have already collapsed and virialized, while others are still in an assembly phase. There are three main modes of mass accretion onto  clusters: the steady infall of matter from the surrounding filamentary large-scale structures, the discrete accretion of group-sized objects, and the extreme event of a major cluster–cluster merger. The latter is the most energetic event known in the universe \citep{Markevitch1998} and results in the violent reassembly of the cluster. Such a dramatic reconfiguration of the cluster results in a rapid change in the environment of its member galaxies. The detailed effect mergers have on galaxies though is yet to be fully understood, given the complex nature of cluster mergers and, hence, the difficulty in obtaining a detailed picture of their properties.

A way to improve our understanding of the role of the cluster environment in galaxy evolution is to focus on a specific peculiar system and study in detail its effect on its member galaxies. One of the better characterized structures at intermediate redshift, a critical epoch during which  clusters are in a fast mass- growing phase (a  cluster with a mass of $10^{14}$ M$_\odot$ doubles its mass between redshift 0.6 and redshift 0, \citealt{Poggianti2006}) is Abell2744 (hereafter A2744). A2744 (also known as the Pandora cluster and as AC118, \citealt{CouchNewell1984}) is an X-ray luminous, merging cluster at $z$= 0.3064, with a virial mass of $7.4\times 10^{15} M_\odot$ (its mass within 1.3 Mpc is $\sim 2 \times 10^{15} M_\odot$, \citealt{Jauzac2016}) and a velocity dispersion $\sigma_{cl}$ of $\sim$1500 km/s \citep{Owers2011}. It is in a particularly dynamic state due to its merging history and  distinct components -- both on the plane of the sky and in redshift space --  have been identified using the X-ray and optical spectroscopy \citep{Owers2011}:  two major substructures, the northern core (NC), and the southern minor remnant core (SMRC) within the cluster, plus a region called the central tidal debris (CTD), which is close in projection to the SMRC but exhibits a velocity close to that of the NC, and a region called Northwestern interloper (NWI), characterized by a cold front to the northeast, and an extended region of enhanced X-ray surface brightness to the south (see Fig. 17 in \citealt{Owers2011} for a visual representation of the different components). This configuration can be explained with a scenario of a post-core-passage major merger \citep[see also][]{GirardiMezzetti2000, KempnerDavid2003, Boschin2006} in addition to an interloping minor merger, with the CTD being a region stripped from the NC by an interaction and the NWI a structure that fell into the main cluster from the south or southeast, initially traveling roughly north or northwest and passing the main cluster core off-center to the southwest. This complex cluster configuration might explain the significant blue galaxy excess of 2.2$\pm$0.3 times that of nearby clusters in the same core regions \citep{butcher84, couch87, Owers2011}, as well as a predominance of starburst and post-starburst galaxies \citep{couch87}. 

Given its remarkable features that make it an ideal target to investigate the environmental effects on galaxies, A2744 has recently been the subject of extended campaigns on VLT/MUSE, which have been used to both better characterize the properties of the cluster as a whole, such the total mass distribution and its lensing capabilities (Bergmaini et al. submitted) and to study the spatially resolved properties of the galaxy members \citep{Moretti2022, Werle2022, Bellhouse2022} for characterizing the link between the cluster properties and star formation activity in cluster galaxies. 

In this paper we exploit $JWST$/NIRCam  observations, combined with VLT/MUSE data,  to investigate the properties of the galaxies in  A2744, to look for evidence of environmental signatures in galaxies located in different regions of the cluster. We parameterize the environment in many different ways, to be sensitive to a wide range of mechanisms taking place in the cluster. Among the many galaxy properties, we focus on galaxy colors and morphologies, to make the best use of the  NIRCam photometry. 

After characterizing the overall population of cluster and field galaxies, we focus on the most striking feature uncovered by our study: a population of ``red-excess'' galaxies, i.e. galaxies with normal optical and F115W$-$F150W colors and an excess of up to 0.8 mags in F200W$-$F444W over the standard red sequence.

A standard cosmology with $\Omega_{\rm m}=0.3$ $\Omega_{\Lambda}=0.7$ and H$_0$=70 km s$^{-1}$ Mpc$^{-1}$ and a \cite{Chabrier2003} Initial Mass Function are adopted.

\section{Data observations and reduction} \label{sec:data}

\subsection{Imaging} \label{sec:data_jwst}
We use JWST/NIRCam imaging obtained by the GLASS-JWST program ERS-1324 \citep{Treu2022}, the UNCOVER\footnote{https://www.stsci.edu/jwst/science-execution/program-information.html?id=2561} program GO-2561 \citep{Bezanson2022} and the Director’s Discretionary Time Program
2756, aimed at following up a Supernova discovered in
GLASS-JWST NIRISS imaging. Taken together, these surveys provide  contiguous coverage over  46.5 square arcmin \citep{Paris2023}, thus covering three out of the four components discussed in Sec.\ref{sec:intro}.

GLASS-JWST NIRCam observations were taken in parallel to NIRISS observations of  A2744 on June 28-29 2022 and in parallel to NIRSpec observations of the cluster on November 10, 2022. They  consist of imaging in seven bands: F090W (total exposure time: 11520 seconds), F115W (11520 s.), F150W (6120 s.), F200W (5400 s.), F277W (5400 s.), F356W (6120 s.), and F444W (23400 s.). As the primary spectroscopic target was the A2744 cluster, these parallel images are offset to the North-West. 

The UNCOVER NIRCam observations were taken on November 2-15 2022. They target the center of the  cluster and the immediate surroundings. These images are composed of four pointing and consist of imaging in seven bands: F115W (10823 s.), F150W (10823 s.), F200W (6700 s.), F277W (6700 s.), F356W (6700 s.), F410M (6700 s.) and F444W (8246 s.).

Finally, the DDT program 2756 (PI W. Chen) targeted the cluster core on October 20 and December 6 2022 (UT). The DDT  filter set is the same as GLASS-JWST with the exception of the F090W filter. 

In our analysis, we also include new and archival HST imaging. For the GLASS fields, this includes HST/ACS data in F606W (59530 s.), F775W (23550 s.), and F814W (123920 s), taken as part of DDT program GO-17231 (P.I. Treu). 

The image reduction and calibration, and the methods used to detect sources and measure multi-band photometry in all fields closely follow that of \cite{Merlin2022} and are presented  by \citet{Paris2023}. 

Source detection was carried out on the F444W band, while aperture fluxes were computed in the other bands on the PSF-matched images. We use corrected aperture photometry, computed as explained by \cite{Merlin2022, Paris2023}.

\subsection{VLT/MUSE spectroscopy} \label{sec:data_muse}
We make use of all the available VLT/MUSE spectroscopy in the surrounding of the A2744 cluster. We limit our analysis to the areas covered by MUSE observations, excluding additional spectroscopy available in the surrounding of the cluster to have at our disposal a spectroscopically complete sample of galaxies (see below). 

For the cluster central region, we rely on the data from the MUSE Lensing Cluster GTO program \citep{Bacon2017, Richard2021}. Data  consist of a $2\times$2 mosaic of GTO observations, with a field of view (FoV) of $\sim 2^\prime \times 2^\prime$  centered on R.A. = 00:14:20.952 and decl. =-30:23:53.88 covering the region that includes the southern and central structures but excludes the northern core and interloper.  Four 1 sq. arcmin quadrants were observed for a total of 3.5, 4, 4 and 5 hours respectively, and the center of the cluster was observed for an additional 2 hours. 

Beyond the cluster central part, we make use of MUSE observations of the GLASS-JWST NIRCam fields  obtained on the nights of July 28 and August 20 2022 through the ESO DDT program 109.24EZ.001 (co-PIs Mason, Vanzella, \citealt{Prieto2022}). The data comprise 5 pointings (4 pointings of which are overlapping with NIRCam imaging, one overlapping only with UNCOVER) each with 1 hour exposure time. The raw data are publicly available on the ESO archive\footnote{\url{http://archive.eso.org/wdb/wdb/eso/sched_rep_arc/query?progid=109.24EZ.001}}. The reduction, calibration and source detection methods used for this work are identical to techniques described in previous works \citep{Caminha2017,Caminha2019}. The final catalog is presented in \cite{Bergamini2023_2}).

\section{Galaxy sample} \label{sec:sample}
\begin{figure}
\centering
	\includegraphics[width=0.45\textwidth]{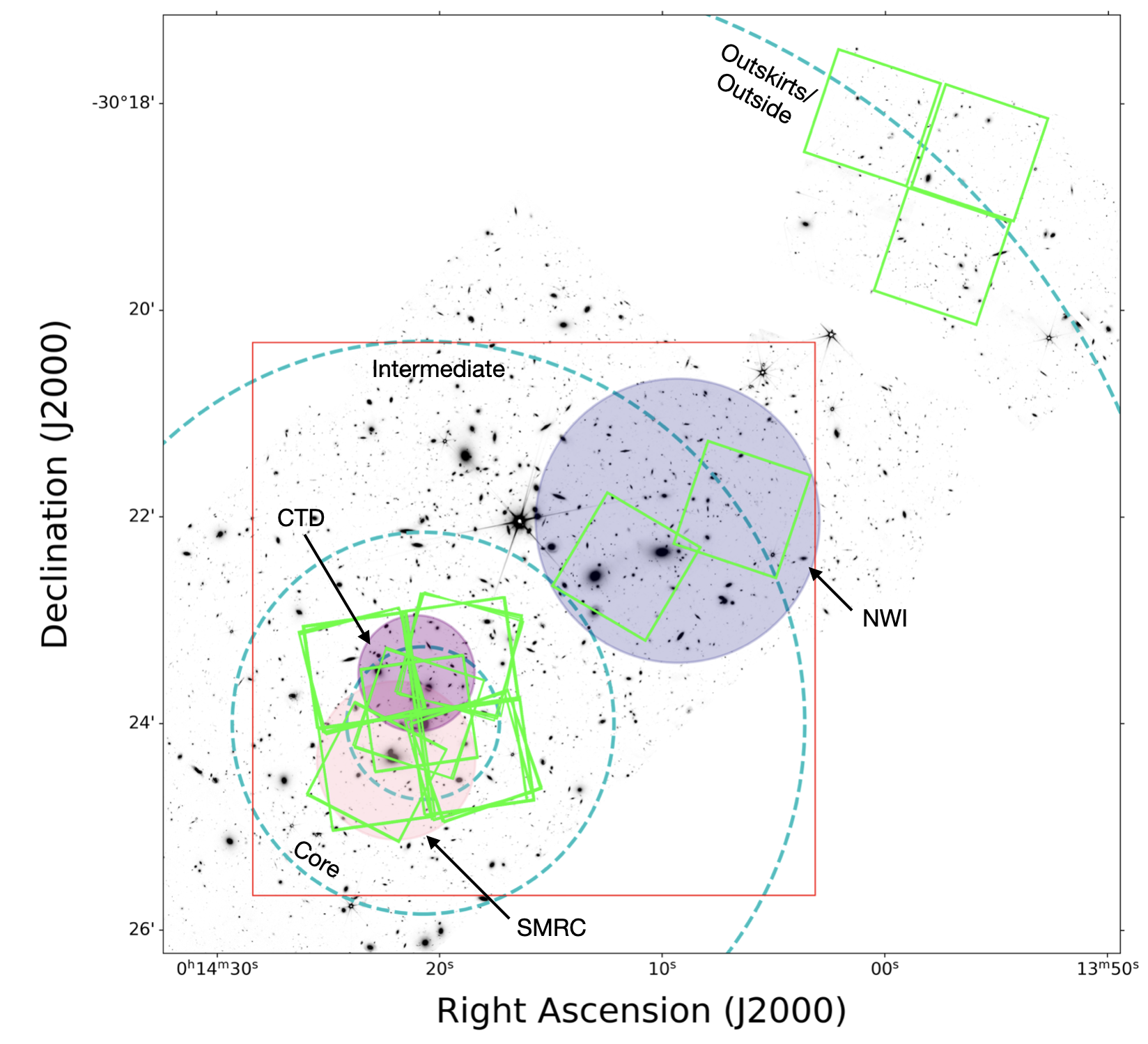}
	\caption{Overview of A2744 showing the NIRCAM F115W image of the cluster, with the MUSE footprint shown by the green squares. Dashed cyan circles indicate regions within 0.1, 0.25, 0.5, 1 $r/r_{200}$, respectively. Filled circles represent approximate location of the SMRC (pink), CTD (purple) and NWI (navy), from \citet{Owers2011}. SMRC and CTD are acutally separated in redshift space, as shown in Fig.\ref{fig:zhist}. The red square represents the area over which the total surface mass density of the cluster has been reconstructed by \citet{Bergamini2023}. }
\label{fig:FoV}
\end{figure}
We select galaxies that are both in the MUSE and NIRCam footprints (Fig.\ref{fig:FoV}). This choice allows us to assemble a sample with a high spectroscopic coverage. Indeed, outside the MUSE pointings, spectroscopically confirmed galaxies are quite sparse. Computing the ratio of the number of spectra yielding a redshift to the total number of galaxies in the F115W catalog across the same regions as a function of F115W magnitude, we set our spectroscopic completeness limit at F115W= 22.2 for the external pointing and 23.3 for the central one ($>$85\% completeness). Throughout the analysis, we will adopt the most conservative value in all regions. 

\begin{figure}
\centering
	\includegraphics[width=0.45\textwidth]{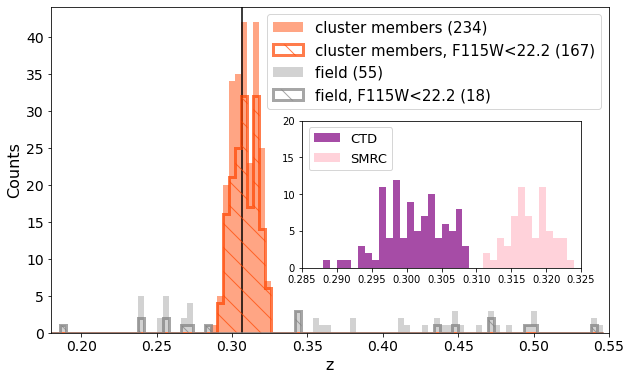}
	\caption{Redshift distribution of the sample. Orange histograms represent cluster members;  grey histograms represent the field sample. Filled histogram show the full sample, hatched histograms galaxies above the magnitude limit. The black vertical line indicates the cluster redshfit. The inset shows the redshift distribution of the SMRC and CTD, highlighting the difference in redshift of the components that instead overlap in space. }
\label{fig:zhist}
\end{figure}

We then separate galaxies based on their global environment: we define as cluster members galaxies   lying within 3$\sigma_{cl}$ from the cluster redshift and assemble a field comparison sample as galaxies in the redshift range $0.15<z<0.287$ and $0.326<z<0.55$.\footnote{Note that the implied stellar mass limits can be up to 1.3 dex different for galaxies at the two edges of the field redshift bin.  Given the small sample statistics, though, we can not adopt the most conservative completeness mass limit.} The final cluster sample, above the magnitude completeness limit,  includes 167 galaxies, the field sample 19. Figure \ref{fig:zhist} shows the redshift distribution for the galaxies in both the cluster and coeval field.

To better quantify the effect of the cluster environment on galaxy properties we adopt four additional different parametrizations. 
\begin{enumerate}
    \item Clustercentric distance. We measure the distance of each galaxy from the cluster core, and scale it by the virial radius. We then separate galaxies into three bins, r=[0, 0.25, 0.5, 1.2] (see Fig.\ref{fig:FoV}). Each bin contains 105 (core), 51 (intermediate), 11 galaxies (outskirts), respectively.  This is one of the most traditional ways to investigate the effect of the environment, but given the merging nature of the cluster, this metric might not be effective in capturing environmental effects. 
    \item Cluster total surface mass density. We use the best-fitting total surface mass density map obtained from an improved version of the strong lensing model presented in \cite{Bergamini2023} (see footprint in Fig.~\ref{fig:FoV}). {We refer to \cite{Bergamini2023} for a more detailed description of the model.} For each galaxy, we subtract its total mass contribution from the originally reconstructed total mass map of the cluster and measure the mean values of the total surface mass density in annuli with radii of 1 and 2 arcsec centered around it. We then consider three equally populated bins of surface mass density. The three bins, whose limits are $\log(\Sigma_{mass}[10^{12}M_\odot/\rm{ kpc^2}]) = [-3.29, -3.02, -2.78, 1.0]$, contain 57, 54, 56 galaxies, respectively.
    \item Projected local galaxy density. {Considering only the spectroscopically confirmed cluster members, above the magnitude completeness limit, for each galaxy} we  compute the circular area containing the five nearest projected neighbours ($A_{5}$) . When this circle extends beyond the field of the observations, we take $A_5$ to be the area of the circle that intersects with the observed field, to correct for edge effects. Local galaxy number densities are then defined as $\rm \Sigma_{5th} = 5/A_5$ in number of galaxies per $\rm Mpc^{2}$.    We then consider three bins of projected local density, $\log(\Sigma_5^{th}[Mpc^2]) = [-3.25, -2.5, -2.12, -1.5]$. These thresholds ensure us to isolate low density regions from the densest one. The different bins contain 25, 71, 71 galaxies, respectively.
    \item Signatures of merger remnants. We select galaxies belonging to the different subcomponents identified by \cite{Owers2011} and discussed in the Introduction (see Fig.\ref{fig:FoV}). Three of these regions are covered by our data. SMRC and CTD are co-spatial, but separated in redshift, as shown in the inset of Fig.\ref{fig:zhist}. 46 galaxies belong to the SMRC, 61 to the CTD (44) 51 to the  NWI. 11 galaxies are outside these regions and correspond to the outskirts described in point 1. above. 
    \end{enumerate}

\begin{figure}
\centering
	\includegraphics[width=0.45\textwidth]{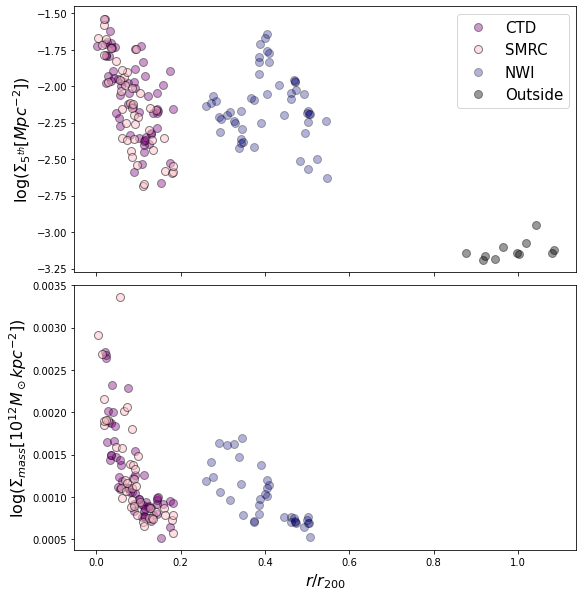}
	\caption{Top: projected local density vs. clustercentric distance. Bottom: projected surface mass density vs. clustercentric distance. Points are colorcoded according to the merger remnant signatures, as indicated in the labels. 
 }
\label{fig:dist}
\end{figure}

Figure \ref{fig:dist} shows how the different parametrizations of environment are related: while overall an anticorrelation exists between both projected local density/surface mass density and distance, a large scatter dominates the trends, suggesting that each parametrization is sensitive to different conditions in the cluster.\footnote{We note that the gaps in clustercentric distance at $\sim 0.25 r/r_{200}$ and between $0.55<r/r_{200}<0.85$ do not correspond to real gaps, but are due to our sampling, as shown in Fig.\ref{fig:FoV}.}

We stress that the uniqueness of our sample relies on the fact that it has a very high (85\%) spectroscopic completeness and can benefit from an unprecedentedly detailed characterization of environment, in addition to JWST/NIRCam observations, although significantly smaller in number than typical literature samples at similar redshift.

\section{Galaxy properties}

\subsection{Observed and rest frame quantities}\label{sec:sedfitting}
Observed magnitudes and colors are described by \citet{Paris2023}. Here we use the total fluxes, measured with \textsc{a-phot} on the detection image F444W by means of a Kron elliptical aperture \citep{Kron1980}. Total fluxes are obtained in the other bands by normalizing the colors in a given aperture to the F444W total flux.

We obtain rest frame colors, stellar masses and attenuation by fitting synthetic stellar templates to the available NIRCam photometry  with \textsc{zphot }\citep{Fontana2000}, following the same procedure described in \cite{Santini2023}. Briefly, we fix the redshift to the spectroscopic redshift of the galaxy and build the stellar library following the assumptions of \cite{Merlin2021}. 
We adopt \cite{bc03} models, with a \cite{Chabrier2003} IMF, 
including nebular emission lines according to \cite{castellano14} and  \cite{Schaerer2009}. We assume delayed exponentially declining Star Formation Histories  (SFH($t$)$\propto (t^2/\tau) \cdot \exp(-t/\tau)$)  with $\tau$ ranging from 0.1 to 7 Gyr.  We let the age range from 10 Myr to the age of the Universe at each galaxy redshift. Metallicity is allowed to be 0.02, 0.2 and 1 times Solar, 
and dust extinction is assumed to follow a \cite{calzetti00} law with E(B-V) varying from 0 to 1.1. 
We compute $1\sigma$  uncertainties on physical properties  by retaining for each object the minimum and maximum fitted masses among all the solutions with a probability $P(\chi^2)>32\%$ of being correct, both fixing the redshift to the best-fit value. 



\subsection{Galaxy morphologies}

We utilize several quantitative measures of the morphologies, based on multiple broadband filters, to map how morphology varies with wavelength. Following \citet{Treu2022_2} (paper XII of this series), for each band, at its original resolution, we derive {five} largely adopted quantitative morphological statistics. Specifically, we focus on concentration, shape asymmetry, clumpiness (also called smootheness), Gini, and M$_{20}$. We refer to the original papers \citep{Abraham2003, Conselice2003, Conselice2014, Lotz2004, Pawlik2016} for a more detailed description of these parameters and to \cite{Treu2022_2} for a detailed description on how quantities were measured on NIRCam data. Briefly, the concentration of light (C)  measure \citep{Conselice2003, Conselice2014} is derived from the ratio of the radii that contain 80\% and 20\% of the total luminosity of a galaxy, giving an indication of the steepness of the light profile of the source. The shape asymmetry ($A_S$) is derived from the difference between the binary detection mask  and the same mask rotated by 180 degrees \citep{Pawlik2016}. $A_S$  measures the morphological asymmetry, regardless of the light distribution inside the galaxy.  
The Gini structural parameter (G) quantifies the degree of inequality of the light distribution in a galaxy \citep{Abraham2003, Lotz2004}, with lower values indicating a more homogeneous distribution and higher values describing a more concentrated source of flux.
M$_{20}$ is defined as the ratio of the second order moment of the brightest 20\% of pixels in a galaxy’s image and the second order moment of the entire image  \citep{Lotz2004}. This parameter is
sensitive to bright features offset from the galaxy center, which makes it suitable for detecting large disturbances in a galaxy’s morphology. 
The clumpiness parameter (cl) quantifies the contribution of small scale structures in a galaxy \citep{Conselice2003}. A completely smooth light distribution in a galaxy without bright small scale structures has cl=0. 
We derive the clumpiness as in  \cite{Calabro2019b}, taking the ratio of the flux in the clumpy regions to the total flux of the galaxy.

We adopt the lines used in \cite{Lotz2008b} to separate different types of galaxies in Gini, M$_{20}$ space, based on the F814W filter:\footnote{For galaxies with no F814W coverage we use the F606W, having checked that classifications are consistent for galaxies with both bands.}
\begin{itemize}
    \item mergers: $G > -0.14 M_{20} + 0.33$; 
    \item E/S0/Sa: $G\leq-0.14 M_{20}+0.33$ and $G>0.14 M_{20} +0.80$;
    \item Sb/Ir: $G\leq -0.14 M_{20} +0.33$ and $G\leq0. 14 M_{20} +0.80$. 
\end{itemize}

\section{Results}
\begin{figure*}
\centering
	\includegraphics[
width=1\textwidth]{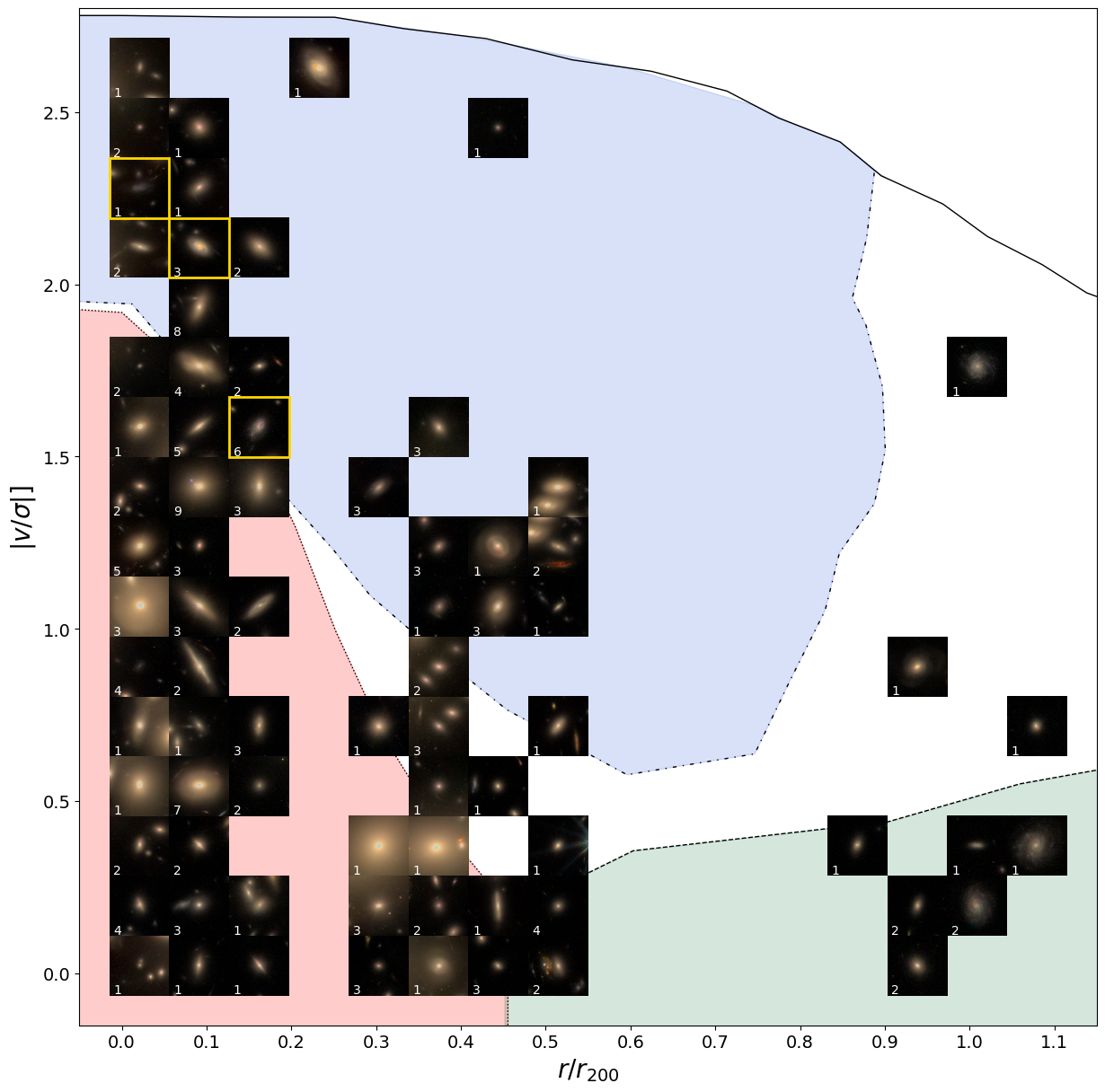}
	\caption{Grid of sample galaxies in the projected phase space. At each location on the grid, an example galaxy is shown in order to visualize the different visual morphologies that occupy different regions of the space. Images are obtained combining the F115W, F150W and F200W passbands. White numbers indicated the number of galaxies in each bin. Galaxies surrounded by a yellow box are cases of ram pressure stripping (see text for details). The different  regions defined by \cite{Rhee2017} are shown with different colors: red = ancient, green = intermediate, blue = recent. The solid black line indicates the limit of subhalos, to define galaxies bounded to the clusters.}
\label{fig:pps}
\end{figure*}

To start, we illustrate the visual morphologies of the galaxies as seen at NIR wavelengths and then we will proceed with a more quantitative analysis of the galaxy colors and morphologies. 

Figure \ref{fig:pps} presents the projected phase space
(galaxy velocity in the cluster versus the clustercentric distance) diagram for the galaxies in the cluster. In the case of relaxed clusters, the position of the galaxies in this space  can be associated with different times since first falling into the cluster. According to \cite{Rhee2017}, four different regions can be identified, where the majority of galaxies lie at a given epoch after they enter in the cluster halo: first (not fallen yet), recent (0 $<t_{\text{infall}}<$3.63 Gyr), intermediate (3.63 $<t_{\text{infall}}<$ 6.45 Gyr), and ancient (6.45 $<t_{\text{infall}}<$ 13.7 Gyr) infallers \citep[see][for other approaches]{Smith2015, Pasquali2019}. Obviously, these numbers need to be taken with caution and galaxies entered at each epoch can be found beyond the corresponding region. As already discussed, A2744 is a complicated structure, and clustercentric distances and relative velocities can not be used to establish time since infall. Nonetheless, it is interesting to inspect morphologies in this plane, for illustrative purposes. The space in Fig. \ref{fig:pps} is divided into bins and an example galaxy, {extracted randomly from the ones entering that bin}, is shown for each bin to indicate typical morphologies corresponding to that combination of parameters.\footnote{We remind the reader  that the observed gaps in clustercentric distance are due to our sampling, as shown in Fig.\ref{fig:FoV}.}

A variety of morphologies is evident and the mix has some dependencies on the position within the plane. Galaxies in the outskirts still present spiral-like morphologies and are all disky, and their morphology seems not to strongly depend on their velocity. Among these ones, it is interesting to note that the position of those with low relative velocities, hence located in the region preferentially occupied by intermediate infallers, supports the idea that these objects have experienced a slow cluster effect for a long time, which has not altered the morphology. 

Galaxies  located in the region of the recent infallers, hence at rather high relative velocities, generally appear to have a spiral morphology, suggesting that the hostile cluster environment has not had the time yet to affect the galaxy appearance. Among these ones, a few galaxies with very high velocities show asymmetries and signs of ram pressure stripping. Some of them were previously known \citep{Owers2012, Moretti2022, Bellhouse2022, Rawle2014}, while a few are new discoveries. We return to this in Sec. \ref{sec:red_gals}. 
In contrast, galaxies with lower velocities present less disturbed disks. 

Galaxies in the region of the ancient infallers present a wide variety of morphologies, from very elliptical cases (especially at very low clustercentric distances) to disky objects with a well defined spiral morphology, similarly to what found by \cite{Kelkar2019}, who analyzed ten clusters between $0.4<z<1$. Undisturbed spiral galaxies hence still survive also in a hostile environment such as the core of a merging system.

\subsection{Galaxy colors}\label{sec:cols}
\begin{figure*}
\centering
	\includegraphics[scale=0.5]{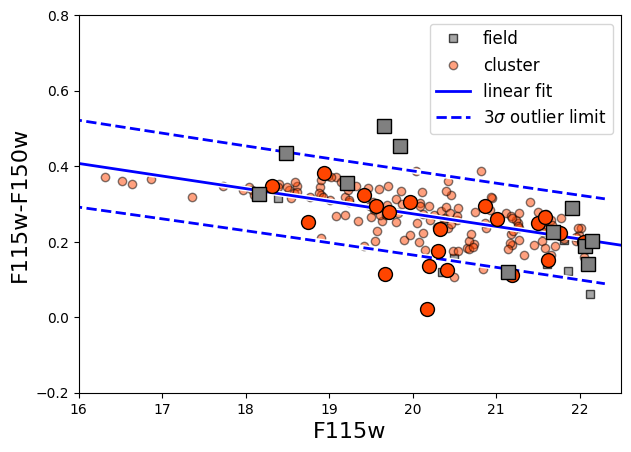}\\
	\includegraphics[scale=0.5]{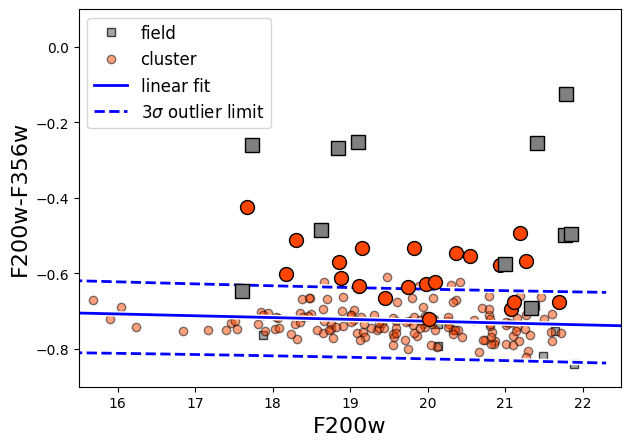}
	\includegraphics[scale=0.5]{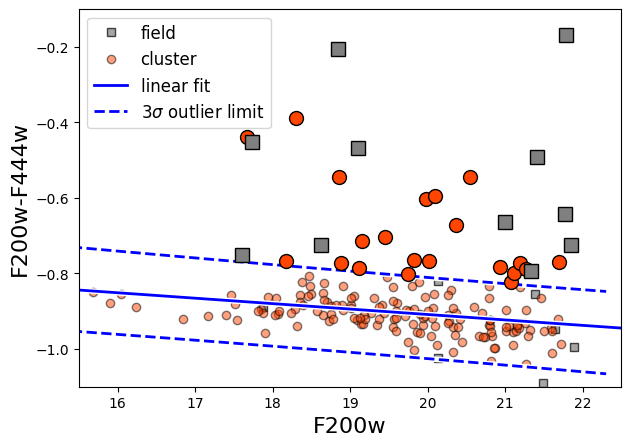}
	\caption{Observed color-magnitude diagram. Upper: F115W$-$F150W vs. F115W. {Bottom right: F200W$-$F356W vs. F200W.} Bottom left: F200W$-$F444W vs. F200W. Red circles: cluster galaxies; grey squares: field galaxies. The blue solid line represents the best fit of the relation, obtained with an iterative 3$\sigma$ clipping procedure. The dashed line shows the $3\sigma$ error. Galaxies  deviating more than $3\sigma$ from the best fit in the F200W$-$F444W vs. F200W plane are indicated with larger symbols in all panels. 
\label{fig:obscolors}}
\end{figure*}

\begin{figure*}
\centering
	\includegraphics[scale=0.4]{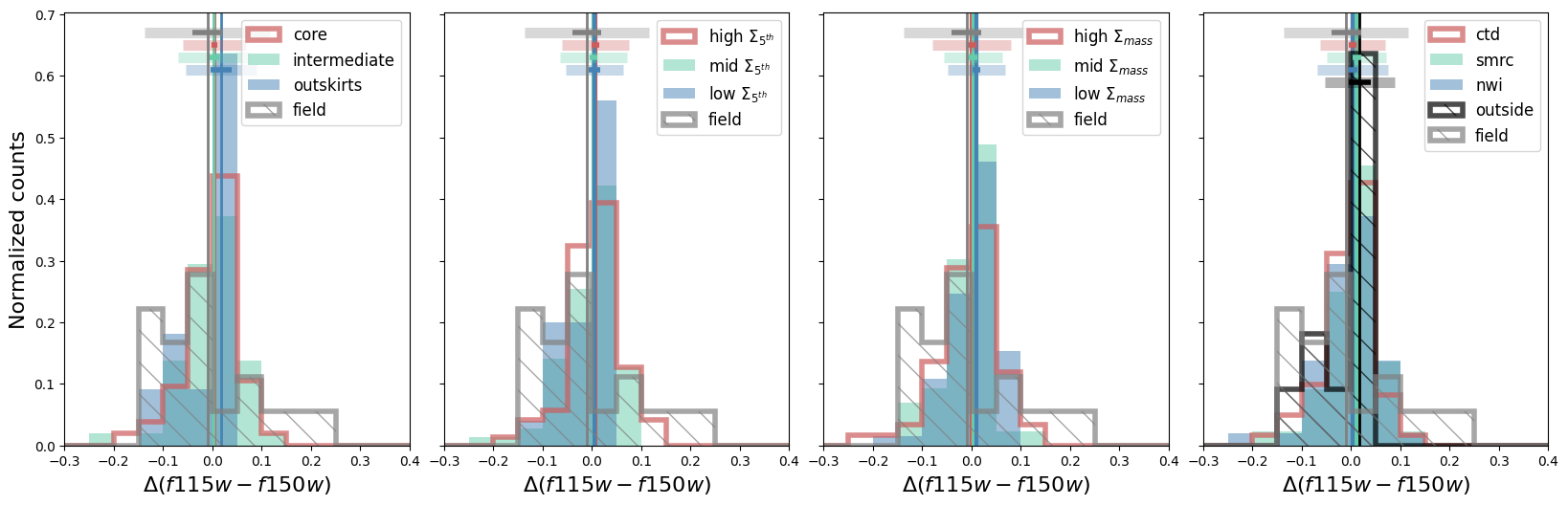}
	\includegraphics[scale=0.4]{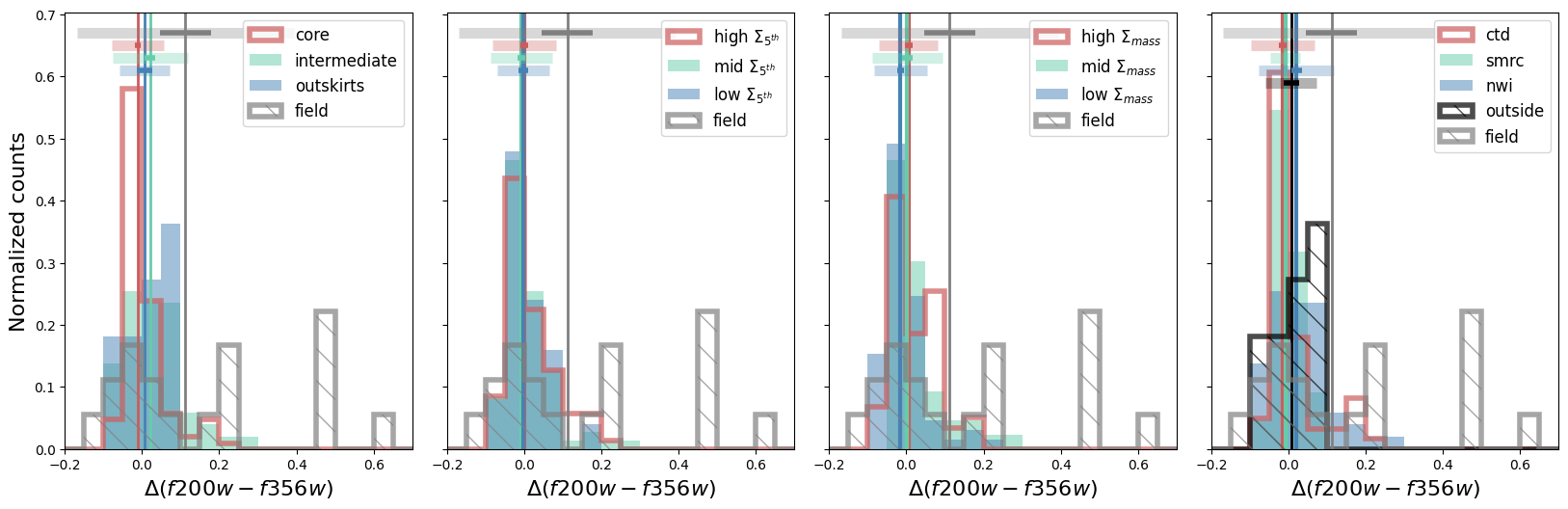}
	\includegraphics[scale=0.4]{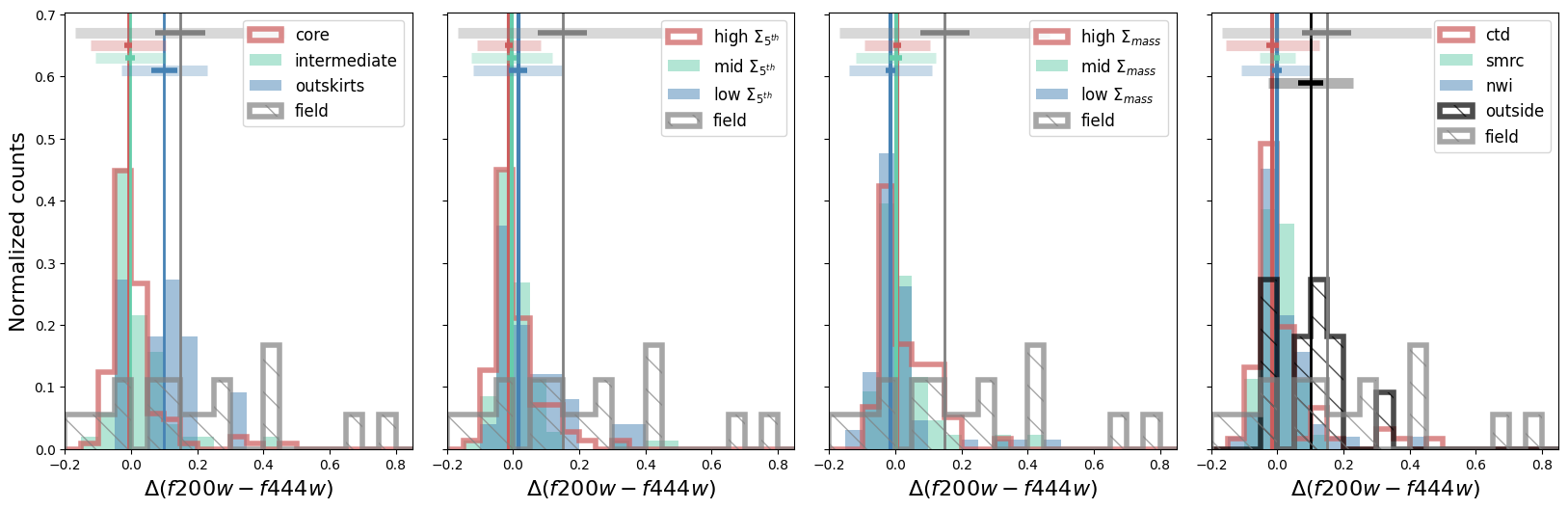}
	\caption{Distributions of the differences between the galaxy color and their expected value according to the fit to entire sample, shown in Fig. \ref{fig:obscolors}, given their magnitude. Upper panels show the F115W$-$F150W color distribution, {central panels show the F200W$-$F356W}, bottom panels show the F200W$-$F444W. From left to right different paramterizations of environment are considered: clustercentric distance, projected local density, surface mass density, signatures of merger remnants. Colors refers to the different subsamples, as indicated in the legend. Vertical lines show median values, thick horizontal lines show the standard error of the distribution ($1.253\times \sigma$, lighter lines) and the error on the median ($1.253\times \sigma/\sqrt{n}$ with $n$ number of galaxies, darker lines).}
\label{fig:delta_obscolors}
\end{figure*}

The new wavelength range covered by JWST allows us to explore a new set of galaxy colors and unveil trends undetected in the optical regime. Figure \ref{fig:obscolors} shows the observed F115W$-$F150W,  {F200W-F356W} and F200W$-$F444W colors, as a function of near-infrared magnitude. The first relation has measured scatter of 0.04 mag. This small value is similar to the scatter of color magnitude relation of local clusters \citep[e.g.][]{Valentinuzzi2011}. 1.6\% of the galaxies lie above 3$\sigma$ from the relation, 2.6\% lie below. No clear differences seem to emerge between cluster and field galaxies.  Considering instead the F200W, {F356W} and F444W bands, {populations of galaxies with significantly red colors emerge.} Fitting the relation and considering galaxies outside the 3$\sigma$ (=0.16 mag) region, {in both cases} we find that  only one field galaxy has a color bluer than the fitted relation, {while 30 galaxies ($\sim$ 16\%)  have significantly redder colors than the bulk of the population in the F200W-F356W color and 32 galaxies ($\sim 17\%$)  in the F200W-F444W colors. In the first case, galaxies can be up to 0.6mag redder than the best-fit, in the second case up to 0.8 mag.} These galaxies include, but are not limited to,  the reddest outliers in the F115W$-$F150W color. Galaxies with normal F115W$-$F150W colors can be much redder at longer wavelengths. {The two populations of galaxies above the best fit relation overall overlap, even though 5 galaxies are outliers only in the F200W-F356W color, 7 only in the F200W-F444W color. The exact overlap though is very sensitive to the threshold adopted to select them. From now on, we will call ``red-excess'' outlier galaxies in the F200W-F444W, the set of bands where deviations are the largest.}
In Sec. \ref{sec:red_gals} we will characterize this population in more detail. 

We proceed by investigating the variation of galaxy colors with environment. To quantify trends, for each galaxy we measure the difference between the observed color  and the value derived from the fit given the galaxy magnitude. Figure~\ref{fig:delta_obscolors} plots the distribution of such differences, considering the different parametrizations of environment discussed in Sec. \ref{sec:sample}. 

When considering the F115W$-$F150W color, no differences emerge either between the field and cluster or between cluster galaxies in different finer environments. A Kolmogorov-Smirnov (KS) test run pairwise on the different samples never rejects the null hypothesis that samples are drawn from the same distribution with high significance ($P_{val} >0.2$). The only difference is that the  field is  characterized by large scatter in color, much larger (by about 3 times) than any of the other samples.  Considering the finer definitions of environments, no significant differences emerge and distributions are rather peaked. 

{When considering the F200W$-$F356W color, the field is significantly different from other samples and the KS test always supports this statement. No significant other differences emerge when inspecting the other samples. }

Considering instead F200W$-$F444W colors, the distributions are much more widely spread, and hints of environmental effects emerge. The field distribution is characterized by an even larger scatter than in F115W$-$F150W, and the KS test finds significant differences ($P_{val} <10^{-5}$) when the field is compared to all samples, except for the outskirts (which correspond to the outside when using the signature of merging remnants) and low density regions. The field median value is almost 0.2 mag redder than the core/high density regions, and median values are statistically different ($>2 \sigma$) even though the large scatter prevents us from drawing solid conclusions. The field is much alike to the outskirts/outside: despite the small number statistics, distributions are peaked at the same value and the KS test establishes with high significance ($P_{val} <0.02$) that the outskirts/outside  are different from both the core and intermediate regions. 
In contrast, galaxies within 0.5$r_{200}$ have a rather peaked distribution centered around 0 and no differences emerge between the two populations. 
No additional significant differences emerge using the other parametrizations of environment, even though redder galaxies tend to lie at low values of projected local density. The KS test confirms a different parent distribution for galaxies in the lowest density bin compared to the other two bins.

\subsubsection{Quiescent fractions in the different environments}
\begin{figure}
\centering
	\includegraphics[width=0.5\textwidth]{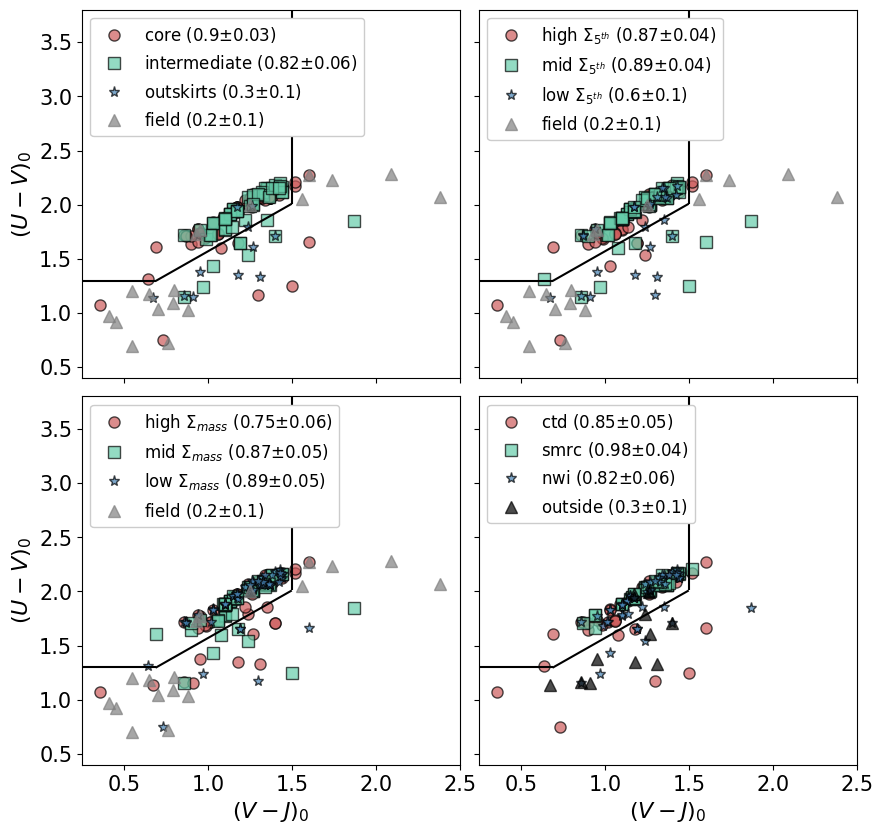}
	\caption{Rest frame U-V vs V-J diagram of galaxies in the different environments. Upper left: clustercentric distance. Upper right: projected local density. Bottom left: surface mass density. Bottom right:  signatures of merger remnants. The separation between star forming and quiescent galaxies is from \citet{Williams2009}. Colors refers to the different subsamples, as indicated in the legend. In the legend, fractions of quiescent galaxies are reported. }
\label{fig:uvj}
\end{figure}

\begin{figure*}
\centering
	\includegraphics[width=0.9\textwidth]{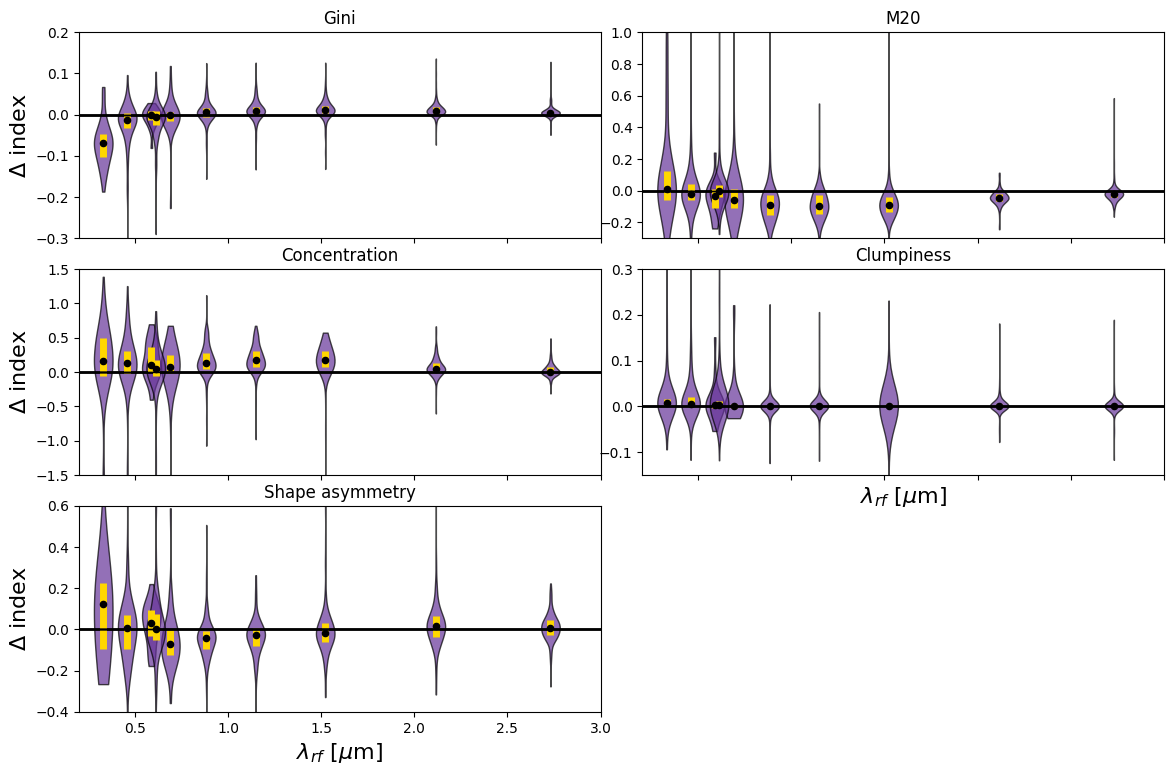}
	\caption{Violin plots distribution of $\Delta$index (= index - index$\rm_{F444W}$) for five morphological parameters as a function of rest-frame wavelength. All galaxies in the sample, regardless of environment, have been considered. Black points represent median values, gold bar the first and third interquartile range. The horizontal black line shows the zero value, which corresponds to no offset. }
\label{fig:pars}
\end{figure*}

In addition to observed colors, we next inspect rest frame colors. Fig. \ref{fig:uvj} shows the rest-frame U-V vs. V-J color diagrams for galaxies in the different environments. Star-forming and quiescent galaxies have been shown to exhibit a bimodality in this plane out to $z\sim 3$ \citep{Labbe2005, Williams2009, Muzzin2013b, Whitaker2011, Straatman2014,  Straatman2016}. 
As expected, the sequence of quiescent galaxies is well visible. The fraction of quiescent galaxies strongly depends on the environment: in the cluster cores it is maximum (90$\pm$3\%), while in the field it is minimum (20$\pm$10\%). Considering cluster outskirts, the fraction of quiescent galaxies is comparable to that of the field, suggesting that the star formation activity is not significantly affected at the virial radius, in agreement with a vast body of literature. 
At 0.5$r_{200}$ the fraction of quiescent galaxies is intermediate between the core and the outskirts. Considering the projected local density, the fraction of quiescent galaxies  increases from the lowest to the highest density bins,  while no statistically significant differences are observed when considering the surface mass density. We remind the reader, though, that the surface mass density measurements are limited to the central part of the cluster, so the mass range spanned is much smaller than the density range probed. Considering the merger remnants signature, the fraction of quiescent galaxies is systematically higher  in the CTD ($\sim85\%$) {SMRC ($\sim98\%$) -- which corresponds to the cluster core -- and NWI  ($\sim82\%$) than in the regions outside any merger remnant.} It appears evident that past merger have induce a higher fraction of passive galaxies.  We will discuss the position of the red-excess galaxies in this plot in Sec. \ref{sec:red_gals}.

Finally, we  also look for trends in colors considering galaxies in the different regions of the projected phase space, but no significant differences emerge in their mean values and scatter.

To summarize, quiescent fractions based on rest frame optical-NIR color strongly depend on environment, whichever definition is adopted. Considering observed colors, JWST unveils a population of extremely red objects, with normal optical colors, whose origin might be linked to the environment. 

{We note that, given our small sample size, we can not isolate the role of stellar mass in affecting quiescent fractions, which is known to play an important role on galaxy evolution \citep[see, e.g.][]{Peng2010, Vulcani2011, ilbert13}. For similar reasons, we will neglect the role of stellar mass also in what follows.}

\subsection{Galaxy morphologies}
We now investigate how the morphological parameters change across wavelength, considering both HST and JWST bands, hence covering the rest frame 0.3-3.5 $\mu m$ at the redshift of the cluster. 
We consider the Gini, M$_{20}$, concentration, clumpiness and shape asymmetry parameters and for each index we measure the difference between the value in a given band and the value measured in the F444W, called $\Delta$index. 
Fig. \ref{fig:pars} shows how the $\Delta$index 
varies as a function of rest frame wavelength. To increase the statistics, we here consider field and cluster galaxies together, having checked that there are no significant environmental dependencies. 
Data distributions are shown in terms of violin plots, which give the probability density of the data at different values, smoothed by a kernel density estimator.  
To summarize the distributions in each band, we also compute the median value of the $\Delta$index in each band, along with the first and the third interquartiles of its distribution. 

While within uncertainties and scatter, morphological indexes do not vary dramatically as a function of wavelength across the entire dynamic range probed by these observations, some trends emerge. The Gini parameter is the smallest when measured at the bluest wavelengths (from HST), but then it is consistent with the values measured in the F444W from $\lambda_{rf}>0.5 \mu m$. $\Delta (M_{20})$ is consistent with 0 for $\lambda_{rf}<0.7 \mu m$, but then it is $<0$ between 1-2.5 $\mu m$. 
F356W measurements are again compatible with those measured in F444W. 
Concentration is systematically larger than in F444W when measured at $\lambda_{rf}<2 \mu m$, while no deviations  are visible  in the clumpiness. As far as shape asymmetry is concerned, there are hints for a larger asymmetry in the bluest bands and a smaller one on the central one, but larger samples will be needed to confirm such trends. 

These results are consistent with the fact that galaxies can appear substantially different at shorter wavelengths than at longer ones \citep[e.g.,][]{Bohlin1991, Kuchinski2000, Windhorst2002}. This shows how little  morphology changes at $\lambda_{rf}>1 \mu m$. Indeed, the optical band light  is a mix of warm and cool stellar components, in which older populations have significant influence. In optical images of spiral galaxies, both the cool bulge and hot spiral arm components can be bright. In contrast, near-infrared light emphasizes cool stellar components, mainly the red giant branch or cool supergiants, while simultaneously reducing the effects of extinction from dust. Spiral structures are often much less prominent \citep{Elmegreen1981}.

\subsubsection{The morphological mix in the different environments}
\begin{figure*}
\centering
	\includegraphics[width=0.4\textwidth]{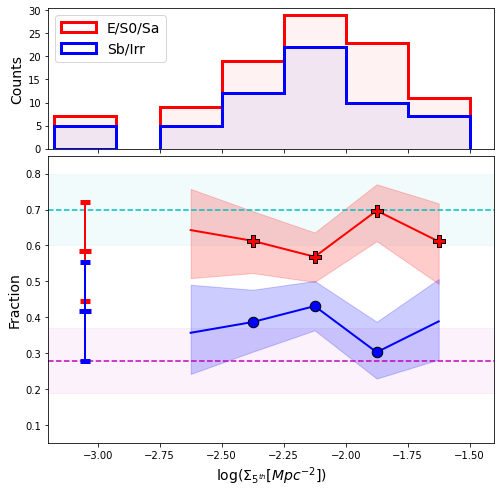}
	\includegraphics[width=0.4\textwidth]{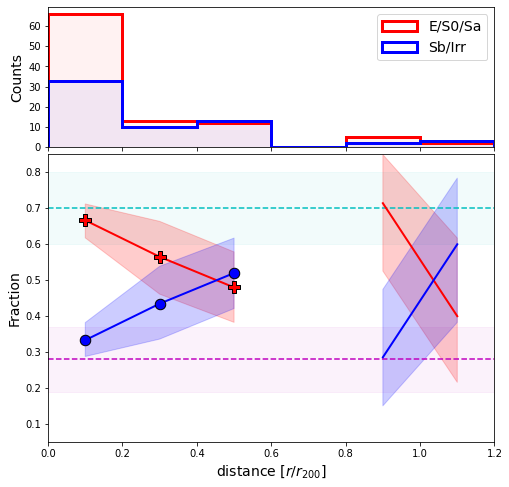}
	\includegraphics[width=0.4\textwidth]{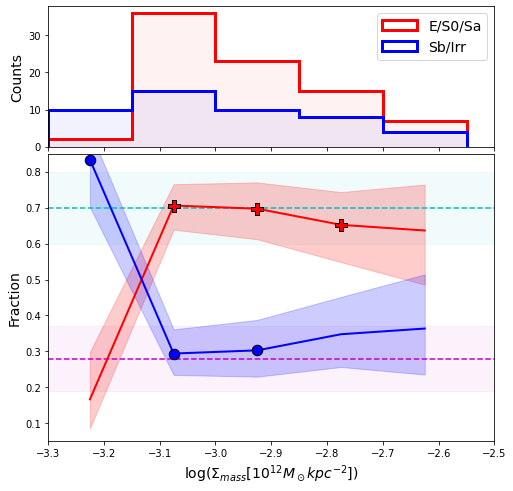}
	\includegraphics[width=0.4\textwidth]{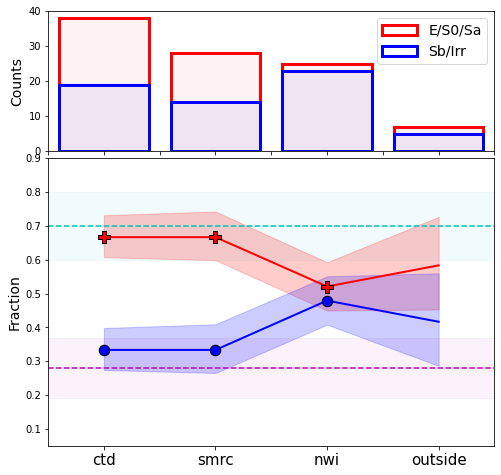}
	\caption{Morphological fractions as a function of the different characterizations of environment. Each quadrant shows a different paramterization: clustercentric distance (top left), projected local density (top right), surface mass density (bottom left), signatures of past merger (bottom right). In each quadrant, the top panel shows the distribution of galaxies as a function on environment, the bottom panel the morphological fraction. Red crosses and lines, and blue circles and lines points represent E/S0/Sa, and Sb/Irr, respectively. Symbols are plotted only for bins with at least 10 galaxies. Shaded areas represent uncertainties, computed as binomial errors. Horizontal lines refer to values and errors measured in the coeval field (cyan for  Sb/Irr, magenta for E/S0/Sa).}
\label{fig:morph_env}
\end{figure*}

Next, using the combination of Gini and M$_{20}$ as discussed in Sec. \ref{sec:sample}, we
obtain an estimate of galaxy morphology. In the cluster, 98 galaxies (0.60$\pm$0.03) are classified as E/S0/Sa, 61 (0.38$\pm$0.04) as Sb/Irr and 3 (0.02$\pm$0.01) as merger. In the field, the number of E/S0/Sa is 5 (0.3$\pm$0.1), of Sb/Irr is 12 (0.7$\pm$0.1), of mergers is 0. In agreement with  previous results \citep[e.g.][]{Dressler1980, Calvi2012, Fasano2015, Vulcani2023}, the effect of the environment is clearly visible on the morphological  mix. 
To better quantify where in the cluster galaxies of different morphologies are preferentially located, Fig. \ref{fig:morph_env} shows the morphology-environment relations, using the four different parametrizations of environment. 

Overall, E/S0/Sa galaxies dominate at all projected local densities, surface mass densities and regardless of merger remnant signatures. They are as common as Sb/Irr only at the lowest densities and outside any merger remnant structure. Only at the lowest surface mass densities Sb/Irr dominate. Considering clustercentric distance, the fraction of E/S0/Sa decreases from the core to the intermediate regions, where the incidence of E/S0/Sa and Sb/Irr is similar.

\section{A F200W-F444W ``red-excess'' population}\label{sec:red_gals}
\begin{table*}
\centering
\caption{Properties of the subsample of galaxies with the F200W$-$F444W excess. \label{tab:gals}}
\begin{tabular}{rllrrrrl}
\hline
  \multicolumn{1}{c}{ID} &
  \multicolumn{1}{c}{RA2000} &
  \multicolumn{1}{c}{DEC2000} &
  \multicolumn{1}{c}{z} &
  \multicolumn{1}{c}{F\_f200W} &
  \multicolumn{1}{c}{F\_f444W} &
  \multicolumn{1}{c}{Env} &
  \multicolumn{1}{c}{dist} \\
  \multicolumn{1}{c}{} &
  \multicolumn{1}{c}{(J2000)} &
  \multicolumn{1}{c}{(J2000)} &
  \multicolumn{1}{c}{}  &
  \multicolumn{1}{c}{(10$^{-6}$ Jy)} &
  \multicolumn{1}{c}{(10$^{-6}$ Jy)} &
  \multicolumn{1}{c}{} &
  \multicolumn{1}{c}{r/r$_{200}$}\\
\hline
  70003396 & 00:13:55.75 & -30:19:54.8 & 0.3122 & 60.38$\pm$0.03 & 31.62$\pm$0.02 & C & 0.92\\
  70006036 & 00:13:56.52 & -30:18:45.9 & 0.3082 & 35.84$\pm$0.04 & 17.69$\pm$0.05 & C & 1.0\\
  70004788 & 00:13:57.13 & -30:19:13.2 & 0.3075 & 13.54$\pm$0.01 & 6.34$\pm$0.01 & C & 0.95\\
  70004550 & 00:13:57.27 & -30:19:27.6 & 0.3072 & 13.0$\pm$0.01 & 6.22$\pm$0.02 & C & 0.92\\
  70003713 & 00:13:59.92 & -30:17:58.3 & 0.3089 & 7.61$\pm$0.02 & 3.74$\pm$0.02 & C & 1.02\\
  70003666 & 00:14:01.24 & -30:17:57.3 & 0.3177 & 36.94$\pm$0.03 & 21.21$\pm$0.04 & C & 1.0\\
  70001043 & 00:14:07.04 & -30:21:53.3 & 0.3067 & 25.9$\pm$0.02 & 13.98$\pm$0.02 & C & 0.49\\
  70003476 & 00:14:11.62 & -30:22:24.5 & 0.2984 & 42.85$\pm$0.07 & 21.2$\pm$0.03 & C & 0.34\\
  70006666 & 00:14:13.71 & -30:22:40.2 & 0.308 & 101.9$\pm$0.05 & 49.98$\pm$0.03 & C & 0.27\\
  70003777 & 00:14:14.08 & -30:22:33.1 & 0.3052 & 12.18$\pm$0.02 & 5.98$\pm$0.02 & C & 0.28\\
  70003981 & 00:14:14.38 & -30:22:40.0 & 0.3234 & 309.56$\pm$0.07 & 206.65$\pm$0.09 & C & 0.26\\
  713 & 00:14:16.63 & -30:23:03.3 & 0.2961 & 33.08$\pm$0.03 & 19.11$\pm$0.04 & C & 0.18\\
  36843 & 00:14:18.98 & -30:24:00.3 & 0.3056 & 15.44$\pm$0.02 & 7.5$\pm$0.03 & C & 0.05\\
  40243 & 00:14:19.43 & -30:23:26.9 & 0.2931 & 104.46$\pm$0.03 & 63.31$\pm$0.03 & C & 0.08\\
  36211 & 00:14:20.14 & -30:24:07.3 & 0.301 & 11.27$\pm$0.04 & 5.45$\pm$0.02 & C & 0.02\\
  35908 & 00:14:20.41 & -30:24:11.9 & 0.3037 & 196.12$\pm$0.1 & 96.82$\pm$0.09 & C & 0.03\\
  35693 & 00:14:20.89 & -30:24:17.8 & 0.2987 & 82.31$\pm$0.04 & 39.9$\pm$0.03 & C & 0.04\\
  35576 & 00:14:21.11 & -30:24:15.3 & 0.2998 & 79.18$\pm$0.07 & 41.04$\pm$0.04 & C & 0.04\\
  34439 & 00:14:21.67 & -30:24:26.6 & 0.3182 & 46.07$\pm$0.03 & 22.03$\pm$0.03 & C & 0.07\\
  41259 & 00:14:22.39 & -30:23:03.8 & 0.2964 & 174.34$\pm$0.04 & 121.89$\pm$0.03 & C & 0.14\\
  41908 & 00:14:25.06 & -30:23:05.9 & 0.2962 & 22.07$\pm$0.01 & 13.37$\pm$0.02 & C & 0.18\\
  70005598 & 00:13:53.08 & -30:18:07.7 & 0.4378 & 7.13$\pm$0.01 & 3.95$\pm$0.03 & F & \\
  70003440 & 00:13:56.37 & -30:19:47.2 & 0.4494 & 128.96$\pm$0.04 & 66.13$\pm$0.02 & F & \\
  70004621 & 00:13:56.90 & -30:19:29.2 & 0.5414 & 9.96$\pm$0.01 & 6.33$\pm$0.01 & F & \\
  70005765 & 00:13:59.29 & -30:19:18.8 & 0.3432 & 105.58$\pm$0.05 & 87.28$\pm$0.04 & F & \\
  70005157 & 00:13:59.32 & -30:19:07.7 & 0.343 & 14.5$\pm$0.03 & 7.86$\pm$0.03 & F & \\
  70005691 & 00:13:59.42 & -30:19:14.3 & 0.3431 & 10.6$\pm$0.03 & 5.1$\pm$0.03 & F & \\
  70004342 & 00:14:10.29 & -30:22:55.7 & 0.4734 & 6.99$\pm$0.01 & 5.98$\pm$0.03 & F & \\
  41371 & 00:14:17.53 & -30:23:07.1 & 0.4988 & 83.02$\pm$0.04 & 53.98$\pm$0.05 & F & \\
  37672 & 00:14:21.58 & -30:23:51.4 & 0.474 & 6.6$\pm$0.02 & 3.39$\pm$0.01 & F & \\
  37257 & 00:14:21.68 & -30:24:01.4 & 0.4971 & 292.16$\pm$0.05 & 192.52$\pm$0.05 & F & \\
  40508 & 00:14:24.01 & -30:23:23.0 & 0.19 & 328.73$\pm$0.04 & 164.78$\pm$0.05 & F & \\
\hline\end{tabular}
\end{table*}

\begin{figure*}
\centering
	\includegraphics[scale=0.6]{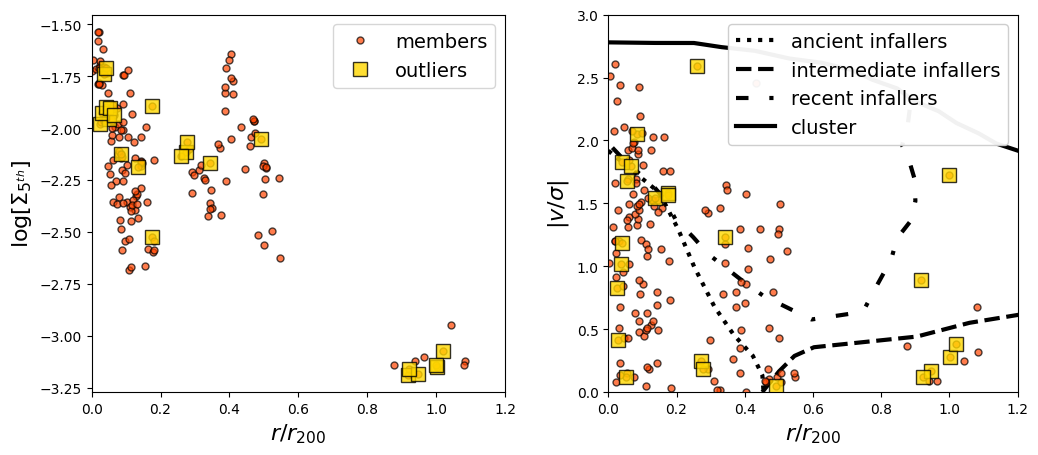}
	\caption{Left: projected local density as a function of clustercentric distance for the cluster population (red circles) and outliers (gold squares). Right: projected phase space diagram for the same two populations. Lines are as in Fig.\ref{fig:pps}.}
\label{fig:out_env}
\end{figure*}

Figure \ref{fig:obscolors} has unveiled an unexpected population of ``red-excess" galaxies, visible only when the reddest JWST bands are considered. 

First of all, to test the reliability of the adopted photometry \cite{Paris2023}, we perform the same analysis discussed in Sec. \ref{sec:cols}, using instead the photometry obtained by \cite{Weaver2023} in the context of UNCOVER, based on completely independent reduction and extraction algorithms. 21/32 galaxies are identified also with the alternative photometry, reassuring us about the reliability of the population. Of the remaining galaxies, 4 have F115W magnitude fainter than our adopted limit of 22.2, according to the \cite{Weaver2023} photometry, 3 have probably have a point-like source that might be affecting the photometry and 3 are on the cluster main sequence according to the UNCOVER photometry. In contrast, 5 additional galaxies are ``red excess" galaxies according to the UNCOVER photometry, one of these is on our red sequence, the others are not in our catalog. 

We now investigate this population in more detail, to understand its origin. In total, 32 galaxies are identified as  outliers, constituting 17$\pm$2\% of the analyzed (cluster+field) population.  Table \ref{tab:gals} summarizes the main properties of the population.

We now characterize various aspects of this population, to shed further light on its origin.

\subsection{The environment of the red excess galaxies}
In Fig.\ref{fig:delta_obscolors} we showed that the observed F200W$-$F444W color of the red excess galaxies, or red outliers,  might be related to the environment, suggesting that the galaxy surrounding might play a role in triggering the observed features. 

21/32 of the red outliers are in the cluster (13$\pm$2\% of the cluster population), 11 in the field (55$\pm$10\% of the field population).
Figure \ref{fig:out_env} shows the location of the galaxies within the cluster.  Half  of the population in the outskirts and about one third of galaxies in low density region have been flagged as red outliers. In such environmental conditions,  we typically do not expect to observe strong ram pressure stripping \citep{Jaffe2018, Gullieuszik2020}, but galaxy galaxy interaction and past minor mergers can occur \citep{Mihos1996}. 
The rest of the red outlier cluster sample is located near the cluster core, but tends to avoid the densest cluster regions and many of them have relatively high velocity. They are in the best position for feeling ram pressure stripping \citep[e.g.,][]{Jaffe2018}. Indeed, some of these galaxies were already identified as potential ram pressure stripping candidates \citep{Owers2012,Rawle2014}. Three of these galaxies have been already confirmed to be  ram pressure stripped galaxies in \cite{Moretti2022, Bellhouse2022}  (A2744\_06, A2744\_09, A2744\_10 in their papers, corresponding to our 40243, 41259, 41908). 
Other two galaxies are outside the area investigated by those papers and two of them were excluded from their analysis due to their red color. 
Of the other galaxies, three (713, 70006036 and 70003981) show signs of stripping, while the others show undisturbed morphology. We discuss morphologies more in detail in the next section.

\subsection{Galaxy morphologies}
63$\pm$10\% of the red outliers are Sb/Irr. They hence are 26$\pm$5\% of the entire Sb/Irr population of the sample analyzed in this work. E/S0/Sa red outliers are only 10$\pm$3\% of the total  E/S0/Sa population.

\begin{figure*}
\centering
	\includegraphics[scale=0.4]{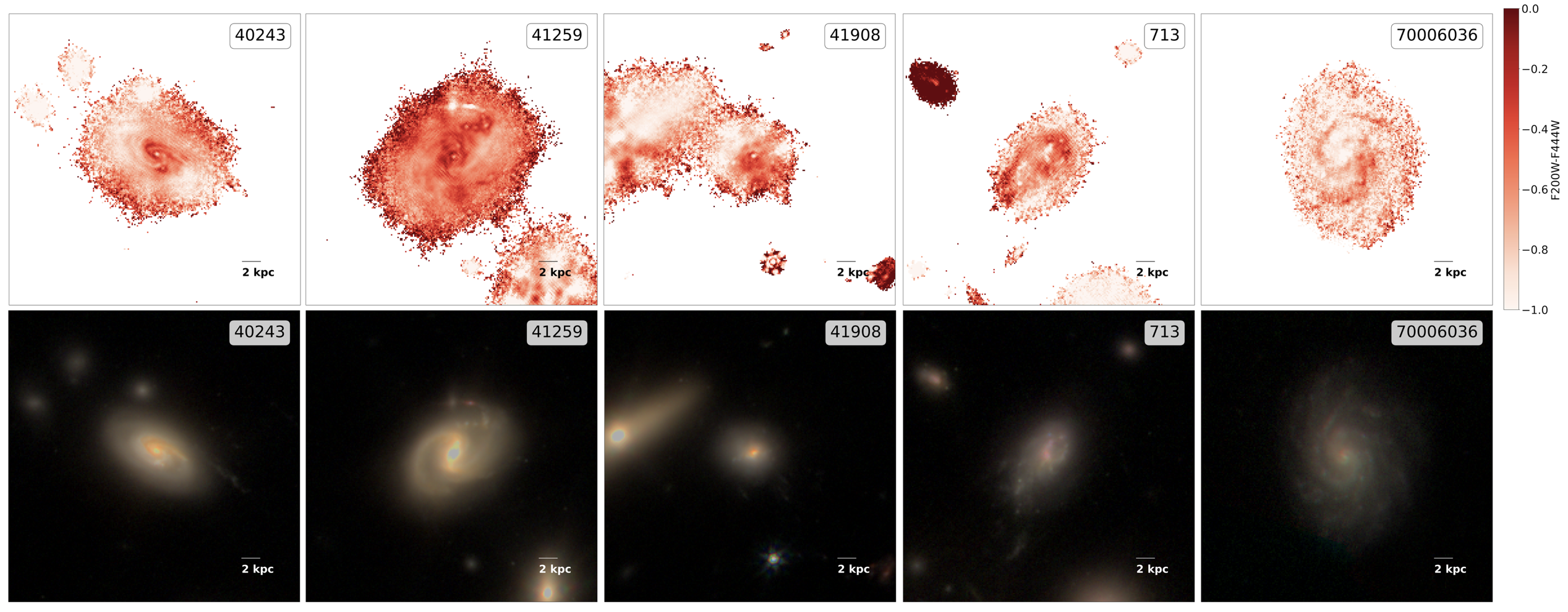}
	\includegraphics[scale=0.4]{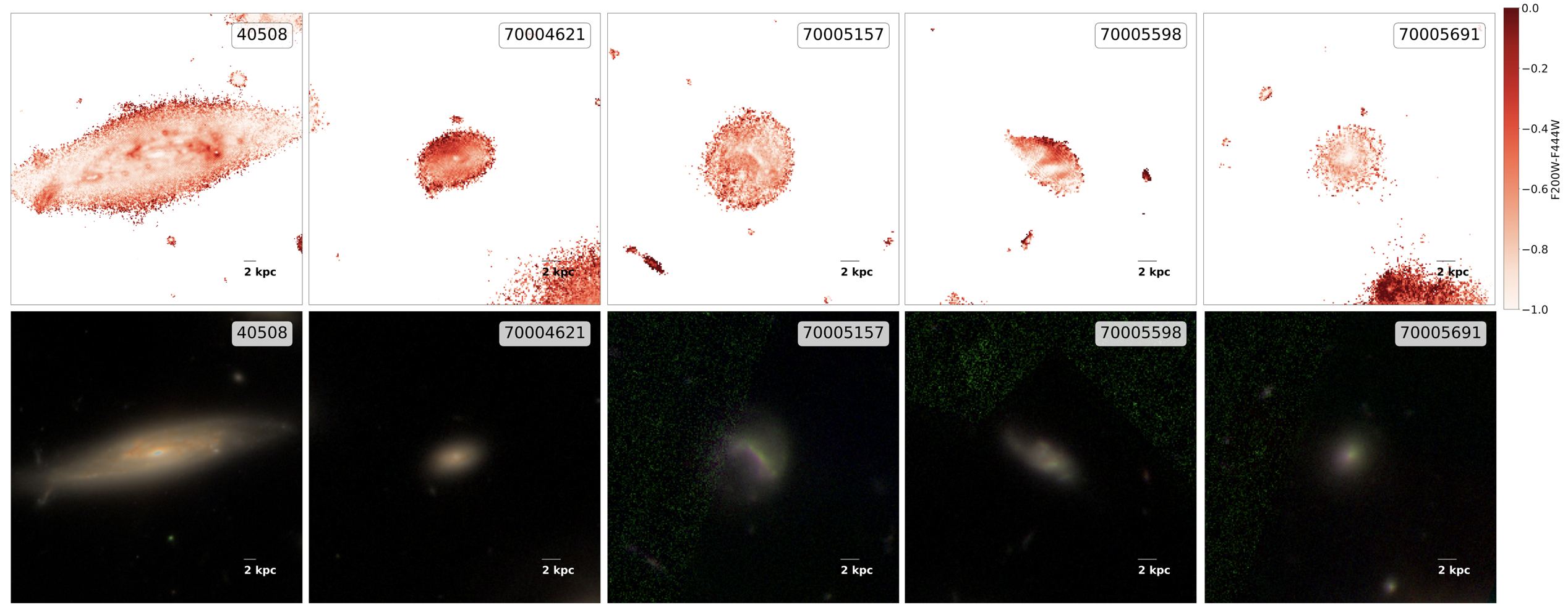}
	\caption{F200W$-$F444W maps and corresponding color composite (F115W+F150W+F200W) images of a subset of the extremely red population identified in Fig.\ref{fig:obscolors}. The top rows shows galaxies in the cluster, the bottom rows show galaxies in the field. The physical scale is reported in the bottom right corners.}
\label{fig:outliers_rgb}
\end{figure*}

Considering the various indicators of galaxy morphology, red outliers have higher values of shape asymmetry and M$_{20}$: the median value of $A_S$ for the red outliers is 0.094$\pm$0.009, while for the normal (cluster+field) population is 0.075$\pm$0.004; a KS test rejects with high confidence ($P_{val} =$0.002)  the hypothesis samples are drawn from the same parent distribution. The  median value of M$_{20}$ for the red outliers is $-$1.98$\pm$0.02, while for the normal (cluster+field) population is $-$1.90$\pm$0.005; the KS test shows this difference is only marginally significant. No differences are detected when comparing median values and distributions of the other parameters. Red outliers hence tend to be more asymmetric and have more bright off center features with respect to the normal population. Disturbances in morphologies must be due to different physical mechanisms in the different environments, as in the field merger and interactions might be responsible for the observed morphology, while in clusters we suggest hydrodynamical mechanisms, as already mentioned. 

To better visualize galaxy morphologies, Figure \ref{fig:outliers_rgb} shows the color composite images for a subset of the red outlier galaxies and the corresponding F200W$-$F444W color map. 
A variety of different morphologies emerge, from smooth to clumpy light distributions, to well defined spiral arms, to disturbed galaxies. Many galaxies show asymmetries, some of them show signs of unwinding arms, similar to what observed in local cluster galaxies \citep{Bellhouse2021, Vulcani2022}.  

The reported cluster galaxies are the four already known ram pressure stripped objects \citep{Owers2012, Moretti2022, Rawle2014} and one of the newly discovered cases. The color map shows that red color excesses are spread throughout the galaxy disks and, when spiral arms are present, they follow them quite accurately. In 41908, the side of the galaxy corresponding to the galaxy tail to the south west is much redder than the other side. 

Also in the field the color excess tightly follows the spiral arms in the undisturbed galaxy (40508). 70005598 is most likely a case of interacting galaxy and a strong gradient in red color excess is observed from north to south, suggesting an environmental influence on this map. In 70004621 and 70005691, two cases of rather small, smooth galaxies, also the color excess is quite smooth.

\subsection{Rest frame properties}
\begin{figure*}
\centering
	\includegraphics[scale=0.5]{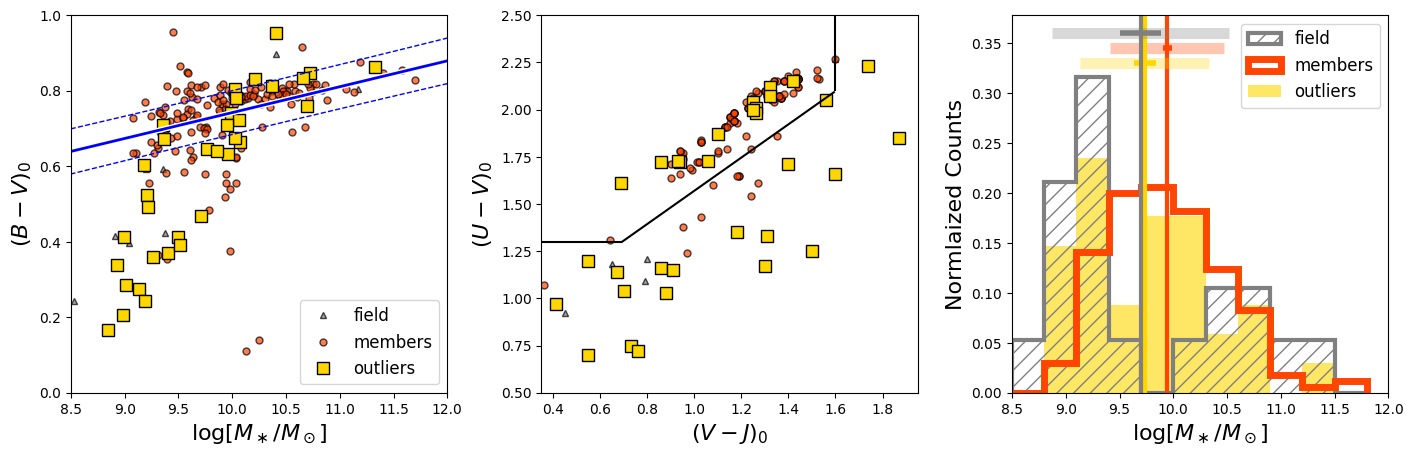}
	\includegraphics[scale=0.5]{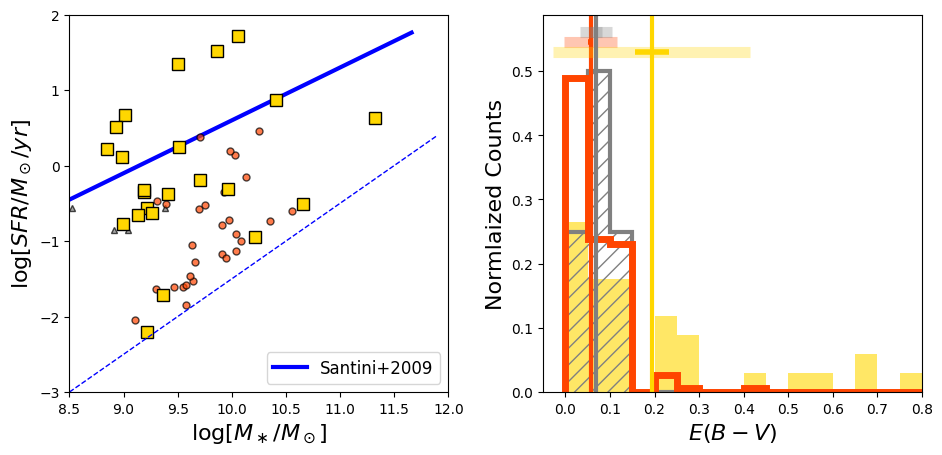}
	\caption{Top left: Rest frame B-V vs stellar mass for the normal cluster members (red circles), normal field galaxies (grey triangles) and the red outliers  (gold squares) identified in Sec. \ref{sec:cols}. The blue line represents the best fit of the red sequence, along with the 3$\sigma$ scatter. Top center: Rest frame U-V vs. V-J diagrams of the outliers, compared to normal field and cluster galaxies. Top right: stellar mass distribution for the different samples. Vertical lines show mean values, thick horizontal lines show the standard error of the distribution (
    lighter lines) and the error on the mean. 
    Bottom left: SFR-mass relation for star forming galaxies. The limit in SSFR (SSFR$>10^{-11.5}$/yr) is shown by the dashed blue line. The SFR-mass relation from \citet{Santini2009} in the redshift range 0.3$<z<$0.6 is also plotted. Bottom right: E(B-V) distribution for the different samples. Vertical lines show mean values, thick horizontal lines show the standard error of the distribution (
    lighter lines) and the error on the mean.
 }
\label{fig:out_colors}
\end{figure*}

We can use the output from the SED fitting (Sec.\ref{sec:sedfitting}) to further characterize the population. 
Figure \ref{fig:out_colors})  shows the location of the red outliers and of the normal population in the rest-frame B-V color vs stellar mass plane, in the UVJ diagram and in the SFR vs mass plane. It also shows the stellar mass and E(B-V) distributions of the different populations. 

Red outliers span a wide range in galaxy colors and stellar mass. 20/32 galaxies lie below the red sequence, defined fitting the entire cluster relation adopting a $3\sigma$ clipping method. Eleven lie on the main sequence, and only one above. Therefore, not all of them are clearly star forming. This is confirmed by the analysis of the U, V, J colors: according to the UVJ diagram, 34$\pm$8\% of the outliers are quiescent. A similar value (29$\pm$8\%) is obtained when using a cut in specific SFR (SSFR) to separate star forming and quiescent galaxies
(SSFR$>10^{-11.5}$/yr). Five of the eleven quiescent red outliers might include a point like source, four of them have a companion. This population should then be taken with caution. 
Overall, red outliers represent 50.0$\pm0.7$\% of the entire (cluster+field) star-forming population. {Considering the star forming galaxies with late type morphology in the cluster core, 6 out of seven are considered red outliers.}

In the cluster, red outliers are typically  less massive than their normal counterparts:  the median stellar mass value of the non outlier cluster population is $\log M_\ast/M_\sun = 10.08\pm 0.05$, that of the cluster outliers is $\log M_\ast/M_\sun = 9.9\pm 0.1$. A KS test  rejects the null hypothesis that the two samples are drawn from the same parent distribution with high confidence ($P_{val}=0.003$).
No differences are found in the field, most likely due to the small sample size. 

The red outliers span a wide range in star forming properties: using the SFR estimated from the SED fitting, the bottom left panel of Fig. \ref{fig:out_colors} shows that while some red outliers are just above the SSFR threshold adopted to separate star forming from quiescent objects, 
many other can reach star formation rates of 100 $M_\odot/yr$. The sample is too small to obtain robust fits to the relation, but clear differences between red outliers and normal galaxies are evident. For reference, we report the fit to the SFR-mass relation obtained by \cite{Santini2009} in the redshift range $0.3<z<0.6$, converted to our adopted IMF.  The work of \cite{Santini2009} is also based on SED-fitting and they used the same code as the one adopted in this work. We note though that the quality of the data, the redshift range and methodologies are different, so results can not be directly compared.  

Red outliers have also systematically higher values of E(B-V) than normal galaxies (bottom right panel of Fig. \ref{fig:out_colors}): the mean E(B-V) of the red outliers is 0.19$\pm$0.02, that of the normal and field cluster galaxies $\sim$ 0.06$\pm$0.02.

The red excess population it is therefore characterized by high values of dust attenuation and SFR, suggesting the presence of dust-obscured star formation.

\subsection{What is the origin of the red excess?}
\begin{figure*}
\centering
	\includegraphics[scale=0.6]{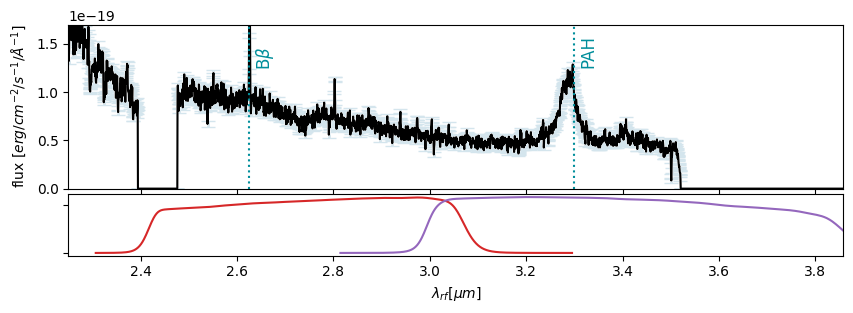}
	\caption{JWST/NIRSpec rest-frame spectrum for 41908, one of the galaxies in our sample characterized by extremely red colors. The position of the Brackett$\beta$ and of the PAH 3.3$\mu m$ is shown. There is no 3.05$\mu$m water absorption though which is seen in spectra of some of the most dust-obscured galaxies. In the bottom panel, the corresponding bandpasses of the broadband F356W and F444W imaging filters are shown.}
\label{fig:spectrum}
\end{figure*}

As part of the GLASS-JWST survey (ERS 1324, PI Treu; \citealt{Treu2022}), we observed one red excess galaxy (41908) with JWST/NIRSpec in multi-object spectroscopy (MOS). We refer to \cite{morishita2022} for details of the NIRSpec observations and data reduction.  Figure \ref{fig:spectrum} shows this spectrum shifted to the rest frame. Two lines are clearly visible: the most prominent feature is the 3.3$\mu m$ Polycyclic Aromatic Hydrocarbon (PAH) line, which has a rest-frame Equivalent Width (EW) of 590$\pm$30 \AA{}. Also the Brackett$\beta$ line is observed, with an emission EW of 30$\pm$1\AA{}. The prominence of these lines can be responsible of the elevated fluxes measured in F356W and especially in F444W. The 3.3$\mu m$ PAH  emission can place constraints on the contribution of dust obscured star formation and active galactic nuclei (AGN) to the cosmic IR background \citep{Schweitzer2006, Spoon2007, Teplitz2007, Valiante2007, Yan2007}.
The rest-frame equivalent width of its 3.3$\mu m$ PAH line puts 41908  exactly at the boundary between ``starburst-dominated" and ``AGN-dominated" galaxy spectra \citep{Inami2018}.
In general, PAH emission features are proposed to be an excellent indicator of star-formation activity \citep[e.g.,][]{Peeters2004, Brandl2006}. PAHs are illuminated by UV photons, mostly from hot stars in star forming regions \citep{Spoon2004, Boselli2004, Calzetti2011}, while they are destroyed by hard radiation from an AGN central engine \citep{Voit1992}.

In our sample, only two of the outliers have broad lines indicative of AGN (70003981 and 70005765) in the MUSE spectra, while none of the other galaxies lie in the region occupied by AGN in the BPT \citep{Baldwin1981} diagram.
Furthermore, another argument against an AGN origin is the fact that we found spatially extended F444W excesses, whereas the hot dust continuum in an AGN would only be present in the central unresolved nucleus.
Thus, we hypothesize that the PAH observed in 41908 is due to heavily obscured star formation, as also suggested by the analysis of the SFR-mass relation and E(B-V) distribution. A larger sample of  NIR spectra will be needed to study in detail the origin of the red-excess.

\begin{figure*}
\centering
	\includegraphics[scale=0.55]{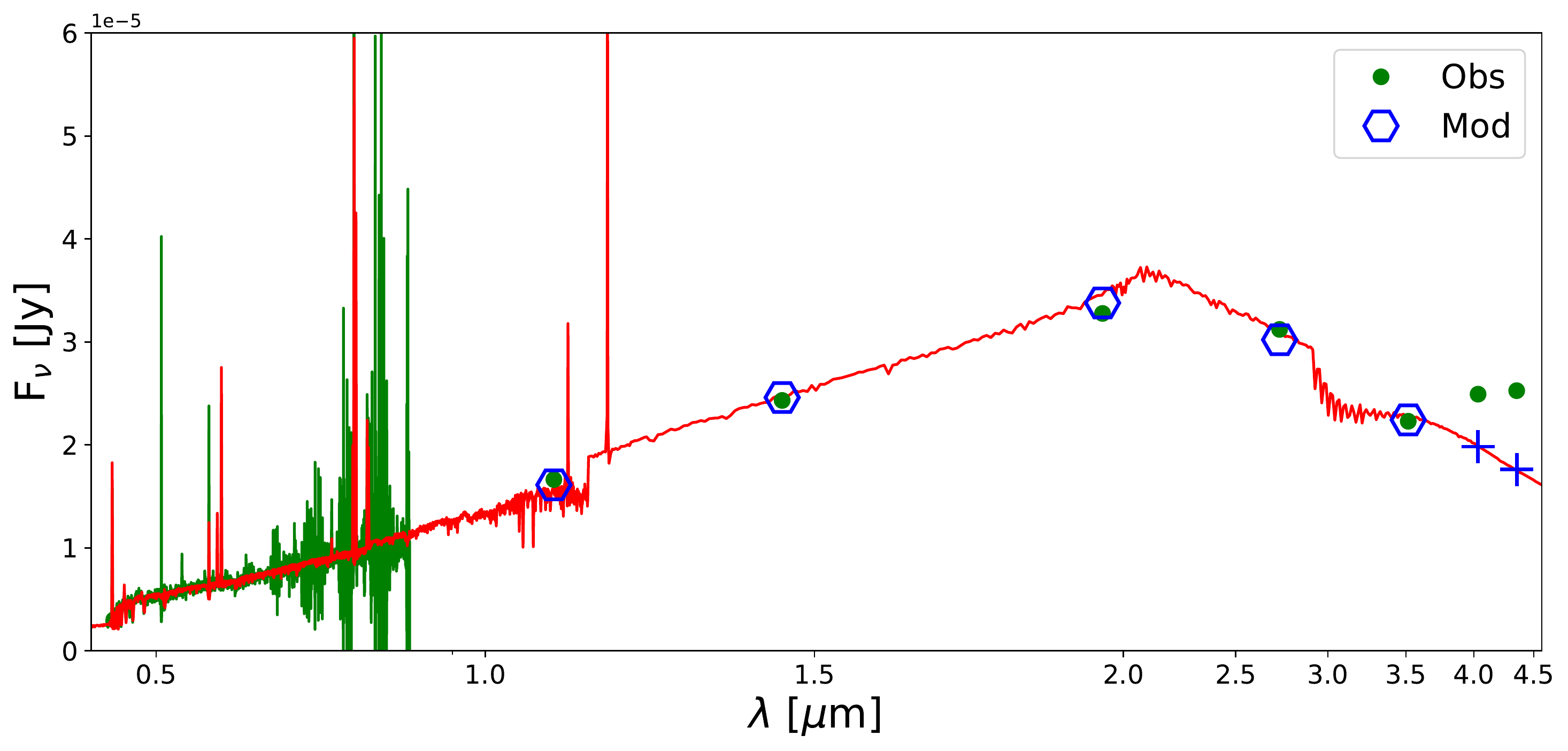}
	\caption{Example of the fitting obtained with the spectrophotomteric modelling. The green line shows the observed MUSE spectrum, the green points the observed JWST/NIRCam photometry. Blue symbols show the values of the flux estimated by the model. Hexagons refer to the bands used in the modelling, crosses to the bands excluded from the modelling. The red line shows the best fit. 
 }
\label{fig:fit}
\end{figure*}

To quantify the amount of the red excess and test if indeed the emission from stellar population is not enough to explain the observed flux, we combine the MUSE spectra with the JWST/NIRCam photometry, both extracted over the same aperture of 2.24$\arcsec$, and fit the full spectral energy distribution with a set templates that  include the emission only from the stellar populations, and not from the PAH.  

We exploit the spectral synthesis code SImulatiNg OPtical Spectra wIth Stellar populations models \citep[SINOPSIS;][]{Fritz2007b, Fritz2011, Fritz2017}. SINOPSIS performs a nonparametric decomposition of galaxy spectra into a combination of single stellar population (SSP) models. 
The fit is performed by minimizing the difference between the model spectrum and the observed spectroscopic data, which include fluxes in selected continuum bands, and the equivalent widths of both absorption and emission lines (the latter consisting of Balmer lines and the {\sc [Oii] 3727 \AA} forbidden line, which are included in the SSP models) \citep[see][for a more detailed description]{Fritz2017}. We also include JWST photometry as a constraint, and the model values are calculated by convolving the spectral model with the appropriate filter response curves. As we are interested in estimating the expected flux at F444W, under the assumption that all the observed emission comes from the pure stellar emission, we exclude this  and (when available) the F410M band from the fitting. 
SINOPSIS combines SSPs of 11 different ages, with the oldest SSP age chosen to be consistent with a galaxy formation redshift of 20.  We assume a constant value of the metallicity as a function of the stellar age, and 7 different values were explored, from Z=0.001 to 0.04. 
Figure \ref{fig:fit} shows one example of the observed data and the best fit obtained by SINOPSIS. 

\begin{figure}
\centering
	\includegraphics[scale=0.6]{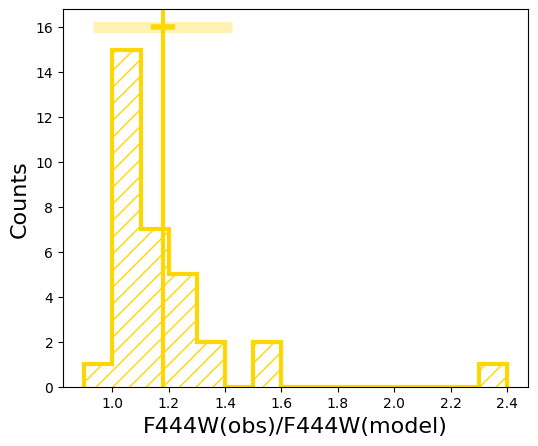}
	\caption{Ratio between the observed flux in F444W and the value obtained by the model for the red outliers. The vertical line shows the mean value, along with the standard deviation and the error on the mean.
 }
\label{fig:ratio}
\end{figure}

From the best fit model, we measure the expected F444W flux and compare it to the observed flux. The distribution of the observed to expected (model) ratios is reported in Fig. \ref{fig:ratio}.
In all but one case the observed flux is indeed larger than what predicted by the model. Typically, the observed value is 20\% larger than the expected one, with one case where it more than twice the expected value. 

This analysis supports our working hypothesis based on only one NIRSpec spectrum that the observed excess is produced by warm/hot dust associated with intense star formation and there is evidence for that in local starbursts \citep{calzetti00}.  The presence of the  3.3 $\mu$m PAH, typically observed in actively star forming galaxies, may play a significant role, even though its equivalent width might not be high enough to lift up the entire broadband F444W filter flux of the observed values ($\sim 20\%$).  The typical rest-frame 
equivalent width of the 3.3 $\mu$m emission in star-forming galaxies is 0.1 $\mu$m. Thus at a redshift of $z=0.3$ it should only brighten the F444W filter by 
about 0.13 mag (and only by 0.20 mag for the most extreme PAH emitters in the GOALS sample,  \citealt{Inami2018}).
In addition, the PAH can not be responsible for the observed boost in F356W and F410W, as its outside the wavelength coverage of those filters. Larger spectroscopic samples targeting the PAH wavelength range are needed to better understand its role.

Alternatively, the excess could also be due to the presence of a deeply obscured AGN, whose presence is not unequivocally revealed by BPT line ratio diagrams. The hot dust emission rising from the dusty torus can contribute to an observed SED approximately at these wavelengths, producing an excess with respect to the pure stellar emission \citep[see e.g.][]{Fritz06}. This scenario would explain particularly well  those objects for which the excess is not only observed in the F444W band, but at also in the F410M one, as there are no known features to be held responsible for the simultaneous excess in the two bands. This hypothesis is though disfavoured by the extent of the red emission, which in case of AGN should be point like. Observations at longer wavelengths (rest frame 5-30 $\mu m$) would be critical to distinguish between these two scenarios.

In principle, none of the proposed scenarios though can fully explain the population of quiescent red outliers, unless they are also extremely obscured star forming galaxies. Only spectra in the wavelength range of the F444W band could explain the origin of such red colors. 

\section{Summary and conclusions}

We combined  JWST/NIRCam imaging and VLT/MUSE data to 
characterize the properties of galaxies in different environmental conditions in the cluster A2744 and in its immediate surroundings. We parameterized the environment in many different ways, to be sensitive to a wide range of mechanisms taking place in the cluster. Thanks to the red wavelength range covered by NIRCam data, we focused on the NIR properties of the galaxies, a so far poorly studied regime. The main results can be summarized as follows:
\begin{itemize}
\item {In the combined cluster and field sample}, structural parameters have little dependence on wavelength: above $\lambda_{rf}>0.9 \mu m$, the median difference between the Gini, M$_{20}$, concentration, clumpiness, shape asymmetry parameters measured in the F444W band in any other band is consistent with zero. Consistent with previous results, morphologies in the NIR bands are rather constants \citep[e.g.][]{Elmegreen1981}. 
\item Galaxies in A2744 are characterized by a varieties of morphologies. Overall, E/S0/Sa dominate at all local densities, distances, surface mass densities and the regions of past mergers. Our sample is too small to determine which of these parametrizations is the main driver. 

\item A non negligible fraction ($\sim17\%$) of galaxies with ``red-excess'' F200W$-$F444W colors (up to  0.8 mag redder than red sequence galaxies) has been found to populate both the cluster regions and the field. {Most of these galaxies are rather red also in the  F200W$-$F356W color, even though the deviation is less pronounced. In contrast, they } have typical F115W$-$F150W colors and rather blue rest frame B-V colors. 
Galaxies with the largest color deviations are found in the field and in the cluster outskirts, suggesting that mechanisms taking place in these regions, might be more effective in producing these colors. These galaxies represent 50\% of the total star forming population, indicating that not all star forming galaxies have such red colors. Many of them occupy the upper envelope of the SFR-mass relation and have high values of E(B-V). Looking at their morphologies, many cluster galaxies have signatures typical of ram pressure stripped galaxies, while field galaxies have features resembling interaction and mergers. 
\end{itemize}

This population of galaxies with the ``red-excess'' resembles various populations of galaxies revealed by the Spitzer Space Telescope with significant excesses in their NIR emission, compared to what is expected from an old stellar population, at various redshifts  \citep[e.g.,][]{Papovich2007, Brand2009, Siana2009, Takagi2010}. At $z\sim 0.3$, though, the PAH at 3.3$\mu m$ was outside the Spitzer window, so previous studies focused mainly on longer wavelengths. \cite{Brand2009} revealed a population of red sequence galaxies with a significant excess in their 24$\mu m$ emission and explained this excess in terms of presence of strong 6.2$\mu m$ and 7.7$\mu m$ PAH emission features in their infrared spectra. They suggested that in a large fraction of the sources the infrared emission is dominated by star-formation processes. Obscured star formation activity may be triggered by minor mergers of red galaxies or, in some cases, may be the residual activity as blue galaxies are quenched to form red sequence galaxies. However, no studies looked for an environmental dependency of this population. 

The  3.3 $\mu m$ PAH feature instead enters the [5.8] IRAC channel at $z>0.6$ and \cite{Magnelli2008} have identified an IRAC excess in the spectral energy distribution of five galaxies, due to the predominance of this feature \citep[see also][]{Mentuch2009, Lange2016}.

The red-excess population also resembles the $z\sim2$ submillimeter-selected galaxies discussed in \cite{Hainline2011}. Fitting the observed-frame optical through mid-IR SEDs, they separated the stellar emission from non-stellar near-IR continuum, finding that about $\sim 10\%$ of their galaxies have significant non-stellar contributions. Since the non-stellar continuum emission is correlated with hard X-ray luminosity,  \cite{Hainline2011} concluded that AGN are at the origin of the emission. 

In line with the interpretation of the IRAC excesses, our hypothesis is that the galaxies discovered here are also characterized by dust enshrouded star formation \cite[see also][]{Ian1999, Duc2002}: we analyzed a NIRSpec spectrum for one of the galaxies and found a strong PAH at 3.3$\mu m$. This feature has been often used to trace dust obscured star formation \citep[e.g.,][]{Magnelli2008, Kim2012}. Also spectrophotometric modeling support this interpretation. On the other hand, this interpretation cannot fully explain the passive red excess galaxies and no scenarios can currently explain them.  Larger spectroscopic samples and spatially resolved spectra are though needed to understand if the color excess is due exclusively to dust-obscured star formation, where in the galaxies the emission comes from,  why not all star forming galaxies are red outliers and why not all red outliers are star forming, and the role of environment in triggering the color excess.

\begin{acknowledgments}
BV thanks the Department of Physics and Astronomy, University of California, Los Angeles, for a very pleasant and productive staying during which most of the work presented in this paper was carried out. We thank the referee for their useful comments.  The data were obtained from the Mikulski Archive for Space Telescopes at the Space Telescope Science Institute, which is operated by the Association of Universities for Research in Astronomy, Inc., under NASA contract NAS 5-03127 for \JWST. These observations are associated with program \JWST-ERS-1324. We acknowledge financial support from NASA through grants \JWST-ERS-1324. BM acknowledges support from an Australian Government Research Training Program (RTP) Scholarship and a Laby PhD Travelling Scholarship.
We acknowledge support from the INAF Large Grant 2022 “Extragalactic Surveys with JWST”  (PI Pentericci). J.F. acknowledges financial support from the UNAM-DGAPA-PAPIIT IN110723 grant, Mexico.
This project has received funding from the European Research Council (ERC) under the Horizon 2020 research and innovation programme (grant agreement N. 833824).
K.G. and T.N. acknowledge support from Australian Research Council Laureate Fellowship FL180100060.
{ All of the data presented in this paper were obtained from the Mikulski Archive for Space Telescopes (MAST) at the Space Telescope Science Institute. The specific observations analyzed can be accessed via \dataset[DOI: 10.17909/9a2g-sj78]{https://doi.org/10.17909/9a2g-sj78} and \dataset[DOI: 10.17909/mrt6-wm89]{https://doi.org/10.17909/mrt6-wm89}.}

\end{acknowledgments}

%

\vspace{5mm}
\facilities{JWST, MUSE/VLT}


\software{astropy \citep{astropy1,astropy2}, grizli \citep{2021zndo...5012699B},
          }






\bibliography{references,references_toadd}{}
\bibliographystyle{aasjournal}



\end{document}